\documentclass[preprint,authoryear, 4p, 12pt]{elsarticle}
\usepackage{color}
\usepackage{lipsum}
\usepackage{amsfonts,amsmath,amssymb,mathrsfs}
\usepackage{graphicx}
\usepackage{epstopdf}
\usepackage[authoryear]{natbib}
\usepackage{algorithmic}
\usepackage{amsopn}
\usepackage{mathtools}
\usepackage{cases}
\usepackage{bbm}
\usepackage{amsmath}
\numberwithin{equation}{section}
\usepackage{enumerate}
\newtheorem{theorem}{Theorem}[section]
\newtheorem{lemma}{Lemma}[section]

\newtheorem{remark}{Remark}[section]
\newtheorem{proposition}{Proposition}[section]
\newtheorem{corollary}{Corollary}[section]

\usepackage{geometry}
\renewcommand{\epsilon}{\varepsilon}
\newcommand{\esssup}{\mathop{\operatorname{ess\,sup}}}

\DeclareMathOperator{\essinf}{ess~inf}

\begin{document}
	\journal{arXiv.org}
\begin{frontmatter}
\title{Pricing principle via Tsallis relative entropy in incomplete market} 
\author[math]{Dejian Tian\corref{correspondingauthor}}
\ead{djtian@cumt.edu.cn}
\address[math]{School of Mathematics, China University of Mining and Technology, Xuzhou, P.R. China}
\cortext[correspondingauthor]{Corresponding author}
\begin{abstract}
	
A pricing principle is introduced for non-attainable $q$-exponential bounded contingent claims in an incomplete Brownian motion market setting.  The buyer evaluates the contingent claim under the ``distorted Radon-Nikodym derivative'' and adjustment by Tsallis relative entropy over a family of equivalent martingale measures.  The pricing principle is proved to be a time consistent and arbitrage-free pricing rule.   More importantly, this pricing principle is found to be closely related to backward stochastic differential equations with generators $f(y)|z|^2$ type. The pricing functional is  compatible with prices for attainable claims.  Except translation invariance,  the pricing principle processes lots of elegant properties such as monotonicity and concavity etc. The pricing functional is showed between minimal martingale measure pricing and conditional certainty equivalent pricing under $q$-exponential utility. The asymptotic behavior of the pricing principle for  ambiguity aversion coefficient is also investigated.

\end{abstract}


\end{frontmatter}

\section{Introduction}

\subsection{Model setup and motivation}
Given a time $T<\infty$. Let  $\left(\Omega,\mathscr{F},\left(\mathscr{F}_t\right)_{0\leq t\leq T},\mathbb{P}\right)$ be a filtered probability space. $\left(\mathscr{F}_t\right)_{0\leq t\leq T}$ is the augmented filtration generated by a $(m+n)$-dimensional standard Brownian motion $\overline{W}=\left(W,W^\perp\right)$, where $W$ and $W^\perp$ are the first $m$ and the last $n$ components. We also assume $\left(\mathscr{F}_t\right)_{0\leq t\leq T}$ satisfies the usual hypotheses, completeness and right-continuity.

Consider a market model with $m$ traded assets
whose price processes evolve in the following simple model,
$$S_t=S_0+\int_0^t\lambda_s ds + W_t, ~0\leq t\leq T,$$
where $S_0\in \mathbf{R}^m$ and  $\lambda$ is a uniformly bounded $\mathbf{R}^m$-valued predictable  process with respect to $\mathcal{F}_t^W=\sigma(W_s, 0\leq s\leq t)$, $t\in[0,T]$. For a general price process, this can be achieved under a suitable assumption on its drift and volatility matrix.

We denote by $\mathcal{Q}$ the set of all probability measures on $(\Omega, \mathcal{F}_T)$, equivalent with respect to $\mathbb{P}$. Define $\mathcal{M}$ as the set of probability measures $\mathbb{Q}\in\mathcal{Q}$ such that $(S_t)_{t\in[0,T]}$ is a local martingale with respect to $\mathbb{Q}$. 

By the predictable martingale representation theorem of Brownian motion,  for any $\mathbb{Q}\in\mathcal{M}$ with density process $D_\cdot^{\mathbb{Q},\mathbb{P}}=\mathbb{E}_\mathbb{P}[\frac{d\mathbb{Q}}{d\mathbb{P}}| \mathcal{F}_\cdot]$, there exists a unique $\mathbf{R}^n$-valued predictable  process $(\alpha^\mathbb{Q}_t)_{t\in[0,T]}$ such that $\int_0^\cdot \alpha^\mathbb{Q}_s \cdot dW_s^\perp$ is a local integral martingale on $[0,T]$ and
\begin{equation}\label{eq:dt}
D_t^{\mathbb{Q},\mathbb{P}}=\mathcal{E}(-\lambda\cdot W+\alpha^\mathbb{Q}\cdot W^\perp)_t,  ~~0\leq t\leq T.	
\end{equation}We also denote $D_{s,t}^{\mathbb{Q},\mathbb{P}}=D_t^{\mathbb{Q},\mathbb{P}}/D_s^{\mathbb{Q},\mathbb{P}}$, $0\leq s\leq t\leq T$.

One of the important problems in mathematical
finance is to price the non-attainable contingent claims in the above incomplete market.
In the incomplete case, a general contingent claim is not feasible to create perfectly replicating portfolios.  From an economic point of view, it means that such a claim will have
an intrinsic risk. Therefore, the market must develop  the arbitrage-free pricing in order to specify the appropriate price for the given contingent claim.

\cite{FS90} propose the
\textit{minimal martingale measure pricing} to solve the above problem.  
The minimal  martingale measure (MEMM) $\mathbb{Q}^{min}$ is defined by
$$\frac{~d\mathbb{Q}^{min}}{d\mathbb{P}}=\mathcal{E}(-\lambda\cdot W)_T=\exp\left (-\int_0^T\lambda_s\cdot dW_s-\frac{1}{2}\int_0^T|\lambda_s|^2ds\right),$$ see Theorem 3.5 in  \cite{FS90}.   In particular, since $\mathbb{Q}^{min}\sim \mathbb{P}$,  for each $\mathbb{Q}\in\mathcal{M}$, then we have  $D^{\mathbb{Q},\mathbb{Q}^{min}}=\mathcal{E}(\alpha^\mathbb{Q}\cdot W^\perp)$.  \cite{FS90}  also show that minimal martingale measure $\mathbb{Q}^{min}$  is related to the relative entropy among all martingale measures.
The relative entropy (also called \textit{Kullback-Leibler divergence}) between any two probability measures $\mathbb{Q}$ and $\mathbb{P}$ is defined by $$H_1(\mathbb{Q}|\mathbb{P}):=\mathbb{E}_{\mathbb{P}}[\frac{d\mathbb{Q}}{d\mathbb{P}}\ln\frac{d\mathbb{Q}}{d\mathbb{P}}]=\mathbb{E}_{\mathbb{P}}[D_{T}^{\mathbb{Q},\mathbb{P}}\ln D_{T}^{\mathbb{Q},\mathbb{P}}],   \quad \textrm{~when~} \mathbb{Q}\ll \mathbb{P}.$$
Relative entropy theory has been proved to have important applications in mathematical finance. For example, the reader can refer to  \cite{F00},  \cite{DG02},  \cite{KS07} and  \cite{FS16}.

The purpose of this paper is to provide new ideas and insights for pricing the non-attainable contingent claim in incomplete market.   We first give the economic motivations  for our problem\footnote{Many thanks to one referee for his/her suggestions on the expression of the motivation. }.   Consider a decision maker (DM) who is concern about model misspecification. The DM has a baseline model $\mathbb{Q}^{min}$,  but also considers alternative models in $\mathcal{M}_f$, which is the set of equivalent  local martingale measures with finite discrepancy regarding to the baseline model and will be given specifically in Section \ref{sec3}.   For some contingent claim $\xi$, the DM evaluates it by the following \textit{pricing principle}
\begin{align}\label{motivation}
F_t(\xi):= \mathop{\essinf}_{\mathbb{Q}\in \mathcal{M}_f} \left( \mathbb{E}_{\mathbb{Q}^{min}}\Big[(D_{t,T}^{\mathbb{Q},\mathbb{Q}^{min}})^q~\xi~\big|~\mathcal{F}_t\Big]+ \frac{1}{ \gamma} H_{q,t}(\mathbb{Q}|\mathbb{Q}^{min})\right),  ~~~~t\in[0,T],
\end{align}where $\gamma>0$ is the ambiguity aversion coefficient,  and $$H_{q,t}(\mathbb{Q}|\mathbb{Q}^{min})=\mathbb{E}_{\mathbb{Q}^{min}}\left[(D_{t,T}^{\mathbb{Q},\mathbb{Q}^{min}})^q \ln_{q}D_{t,T}^{\mathbb{Q},\mathbb{Q}^{min}} ~\Big|~\mathcal{F}_t\right]$$
 is a conditional Tsallis relative entropy characterizing the discrepancy. \textit{Tsallis relative entropy} is put forward by   \cite{Tsallis88}  from the viewpoints of statistical physics (see more details in \cite{Tsallis09}). When $\mathbb{Q}\ll \mathbb{P}$,  it is defined by $$H_q(\mathbb{Q}|\mathbb{P}):=\int (\frac{d\mathbb{Q}}{d\mathbb{P}})^q\ln_q(\frac{d\mathbb{Q}}{d\mathbb{P}})d\mathbb{P}= \mathbb{E}_{\mathbb{P}}\Big[(D_{T}^{\mathbb{Q},\mathbb{P}})^q \ln_q D_{T}^{\mathbb{Q},\mathbb{P}}\Big],$$
where $q>0$, $q\neq1$, and $\ln_q(\cdot)$ denotes the  generalized $q$-logarithm function (see Section \ref{sec2} for a detailed expression).  The parameter $q$ can be viewed as a bias or distortion of the original probability measure.

As a consequence,  for  a given alternative model  $\mathbb{Q}\in\mathcal{M}_f$,  the DM evaluates the expected payoff via $\mathbb{E}_{\mathbb{Q}^{min}}\Big[(D_{t,T}^{\mathbb{Q},\mathbb{Q}^{min}})^q~\xi~\big|~\mathcal{F}_t\Big]$ with a weight $(D_{t,T}^{\mathbb{Q},\mathbb{Q}^{min}})^q$.  When $q\in (0,1)$, the DM put less weight on events with large $D_{t,T}^{\mathbb{Q},\mathbb{Q}^{min}}$; when $q>1$,  the DM put more weight on events with large $D_{t,T}^{\mathbb{Q},\mathbb{Q}^{min}}$.  The DM also penalizes models which are too far away from the baseline model $\mathbb{Q}^{min}$. The discrepancy is measured by conditional  Tsallis relative entropy.  In this setting,   the pricing principle  \eqref{motivation} is a \textit{robustness} evaluation.  In particular, when $q=1$, this is \textit{exactly} the formulation of \cite{HS01}, where $H_{1}(\mathbb{Q}|\mathbb{Q}^{min})$ is the relative entropy.  In this way, our pricing principle can be presented as an extension of \cite{HS01} with the relative entropy replaced by Tsallis relative entropy.         

\subsection{The main contributions of this work}  The pricing principle \eqref{motivation} induced by Tsallis relative entropy seems not to be necessarily time consistent or arbitrage-free, because it is a robustness valuation problem and has the nonlinearity brought by the distortion parameter $q$.   The main contribution of this work is that we identify a connection between this pricing principle and backward stochastic differential equations (BSDEs)  with generators $f(y)|z|^{2}$ type.  This kind of BSDEs  are recently  investigated by Bahlali and his coauthors (see \cite{BEO17,BT18, B20}) and by \cite{ZZF21} for more refined results.

We summarize our main theoretical contributions as follows. First, Theorem \ref{shang1} provides the integral representation for the conditional Tsallis relative entropy between equivalent martingale measure and original reference measure. In addition, we also obtain the representation of the conditional Tsallis relative entropy between equivalent martingale measure and minimal martingale measure.  When $q=1$, these representation results are exactly the corresponding formulations of relative entropy (see  \cite{CS05} or  \cite{Skidas03}).

Second,  and most important,  we find a close connection between  the pricing principle  \eqref{motivation} and a specific quadratic BSDE.  In order to investigate the pricing principle,  we introduce two related  problems of BSDEs (Problem 2 and Problem 3) in Section \ref{sec3}.  Problem 2 is a dual formulation, and it is an optimization problem of BSDEs.  Problem 3 is  a specific quadratic BSDE whose generator  involves $y$ and is in form of $g(t,y,z,z^{\perp})=-\lambda_t \cdot z- f(y)|z^{\perp}|^2$.  After solving the uniqueness solution of this BSDE, we show that both Problem 2 and Problem 3 are equal to the pricing principle induced by Tsallis relative entropy (see Theorem \ref{main}).  To the best of our knowledge,  these results are new, which is a complement and enrichment to the relationship between classical relative entropy and quadratic BSDE.  Theorem \ref{main}  indicates that the pricing principle is a time consistent and arbitrage-free pricing rule, and  it is also compatible with risk neutral pricing principle for the attainable claims.


Finally, we study the properties of the pricing principle. The pricing principle processes lots of elegant properties, such as monotonicity, concavity, time consistency and so on. However, different from the relative entropy or exponential preference's situation, the existence of the distortion parameter $(q\neq 1)$ renders this pricing principle does not satisfy the translation invariance or cash-additivity.  Translation invariance is sometimes criticized by some scholars (for example,  \cite{ER09},  \cite{HWWX21}), and our pricing principle sheds some lights on this direction.

We further investigate conditional certainty equivalent for the contingent claim under $q$-exponential utility,  which is proved to be the robust representation by a family of equivalent measures (maybe not martingale measures),  see Proposition \ref{dual-tsallis-z} .   Proposition \ref{jixian23} states that our pricing principle is between minimal martingale measure pricing and conditional certainty equivalent pricing under $q$-exponential utility, which indicates it is a good choice among the arbitrage-free prices.  Moreover,  we consider the asymptotic behavior of the pricing principle when ambiguity aversion coefficient goes to zero or infinity.  

\subsection{Related literature} We end the introduction with some related literature. Another widely used method for valuation in incomplete markets is indifference valuation.  Especially in exponential form, translation invariance can lead to analytical tractability as well as attractive properties. It is known to all that  expected exponential utility indifference pricing, especially in a dynamic context,  involves a quadratic BSDE and relative entropy theory. See  \cite{REK00},  \cite{MS05}, \cite{BK09}, \cite{LQ03},  \cite{FMS11}, \cite{HL14} or \cite{Hu05}. Quadratic BSDEs have been considered by  \cite{K00},  \cite{BH06,BH08},  \cite{XZ18} and so on.

This paper contributes to the literature on pricing the contingent claims in an incomplete market without using classical relative entropy.  Recently, more and more non-relative entropy models have entered the vision of scholars.   The papers closest to ours are  \cite{AG19},  \cite{MT21} and  \cite{MVX21}.

 \cite{AG19} derives a generalization of model uncertainty framework of  \cite{HS01}, using Tsallis relative entropy, and finds calibrations that match detection error probabilities yield comparable asset pricing implications across models.
 \cite{MT21} consider the generalized entropic risk measures, which is related to conditional certainty equivalent of $q$-exponential utility, see subsection \ref{sec4.3}. 
\cite{MVX21} apply Cressie-Read divergence (a general relative entropy which closes to Tsallis relative entropy) to understand portfolio choice and general
equilibrium asset pricing.

 \cite{F00a} constructs a theory of value based on agent's preference and coherent with the no arbitrage principle.   \cite{L08} offers a convex
monotonic pricing functional for non-attainable bounded contingent claims, defined as the convex conjugate of a generalized entropy penalty functional.
 \cite{HPDR10}, \cite{CHKP16} and  \cite{KXZ17} study the
equilibrium pricing using the translation invariant preferences. \cite{FMM11} and \cite{FMM17} investigate the related stochastic control problem by relative entropy and $\phi$-divergences. \cite{LS14} and \cite{CR20} examine the dynamic risk measures and related BSDEs in the jump situation, and the readers can refer to  \cite{R06},  \cite{J08} and  \cite{DPG10} for the Brownian motion case.

The remainder of this paper is organized as follows. Section \ref{sec2}  introduces the definition of Tsallis relative entropy, and provides its integral representations.  Section \ref{sec3} presents our main results. 
The properties of the pricing principle are investigated in Section \ref{sec4}.  Section \ref{sec6} concludes the paper.  All the proofs are relegated to Appendix.

\section{Tsallis relative entropy and its integral representation}\label{sec2}

Similar to relative entropy, Tsallis relative entropy can also be interpreted as how far away $\mathbb{Q}$ is from the reference $\mathbb{P}$; see more details in \cite{Tsallis09}. We will use Tsallis relative entropy to explore its potential implications in pricing principle. This section
introduces the definition of Tsallis relative entropy and investigates its integral representation.

Let's first introduce the  $q$-exponential function and its inverse, the $q$-logarithm function, which are defined respectively as follows

\[\exp_q(x) :=\begin{cases}
[1+(1-q)x]^{\frac{1}{1-q}},  & x\geq -\frac{1}{1-q}  ~\text{ and } ~0<q<1,\\
[1+(1-q)x]^{\frac{1}{1-q}},  & x< -\frac{1}{1-q}~\text{ and } ~q>1,
\end{cases}\]
and
\[\ln_q(x) :=\begin{cases}
\frac{x^{1-q}-1}{1-q},  &x\geq0 ~\text{ and } ~0<q<1,\\
\frac{x^{1-q}-1}{1-q},  & x>0 ~\text{ and } ~q>1.
\end{cases}\]

We always assume that $q>0$ and $q\neq 1$, and $\exp_{q}(\cdot)$ and $\ln_{q}(\cdot)$ are well defined.  $\textrm{Dom}(\ln_{q})$ and $\textrm{Dom}(\exp_{q})$ represent the domains for $\ln_{q}(\cdot)$ and $\exp_{q}(\cdot)$ respectively.  One can easily get that if $q\rightarrow 1$,  then  $\ln_q(x) \rightarrow \ln(x), x>0$, and $\exp_q(x)\rightarrow \exp(x), x\in \mathbf{R}$.

For any two probability measures $\mathbb{Q}$ and $\mathbb{P}$ on $(\Omega, \mathcal{F}_{T})$,  \cite{Tsallis88, Tsallis09}  introduces the  $q$-generalization of the relative entropy ( also called \textit{Tsallis relative entropy}) as follows:
\begin{equation}\label{tsallis}
H_q(\mathbb{Q}|\mathbb{P}) :=
\begin{cases}
\int (\frac{d\mathbb{Q}}{d\mathbb{P}})^q\ln_q(\frac{d\mathbb{Q}}{d\mathbb{P}})d\mathbb{P}, &\text{ $\mathbb{Q}\ll \mathbb{P}$},\\
+\infty, &\text{ others,}
\end{cases}
\end{equation}
where $q>0$ and $q\neq 1$.

Obviously, $H_q(\mathbb{Q}|\mathbb{P})$ is well-defined when $0<q<1$. For $q>1$, if $\mathbb{Q}\ll \mathbb{P}$, then $\mathbb{Q}(\{\frac{d\mathbb{Q}}{d\mathbb{P}}=0\})=\int_{\{\frac{d\mathbb{Q}}{d\mathbb{P}}=0\}}\frac{d\mathbb{Q}}{d\mathbb{P}}d\mathbb{P}=0.$  Hence, $H_q(\mathbb{Q}|\mathbb{P})=\int (\frac{d\mathbb{Q}}{d\mathbb{P}})^{q-1}\ln_q(\frac{d\mathbb{Q}}{d\mathbb{P}})d\mathbb{Q}$ is also well-defined.

Similarly, if $\mathbb{Q}\ll \mathbb{P}$ on $\mathcal{F}_{T}$,  for each $t\in[0,T]$,  
\begin{align}\label{tiaojianshang}
H_{q,t}(\mathbb{Q}|\mathbb{P}) := \mathbb{E}_{\mathbb{P}}\Big[(D^{\mathbb{Q},\mathbb{P}}_{t,T})^q\ln_q D^{\mathbb{Q},\mathbb{P}}_{t,T}~\big|~\mathcal{F}_{t}\Big],\end{align}
denotes  the conditional Tsallis relative entropy of $\mathbb{Q}$ and $\mathbb{P}$ under $\mathcal{F}_{t}$.

In the following, we are going to establish the integral representations for the (conditional) Tsallis relative entropy. 

\begin{theorem}\label{shang1} Let $q>0$ and $q\neq 1$. For any given $\mathbb{Q}\in \mathcal{M}$, suppose $H_q(\mathbb{Q}|\mathbb{P})<+\infty$, then the following results hold.
	\begin{itemize}
		\item[(i)] For each $t\in [0,T]$, we have
		\begin{align}
		H_{q,t}(\mathbb{Q}|\mathbb{P})&=\frac{1}{(D_{t}^{\mathbb{Q},\mathbb{P}})^{q}}\Big( \mathbb{E}_\mathbb{P}\Big[(D_{T}^{\mathbb{Q},\mathbb{P}})^q\ln_qD_{T}^{\mathbb{Q},\mathbb{P}}~\big|~\mathcal{F}_{t}\Big]-(D_{t}^{\mathbb{Q},\mathbb{P}})^q\ln_qD_{t}^{\mathbb{Q},\mathbb{P}}\Big)\label{hqt1}
		\end{align}
and \begin{align}
		H_{q, t}(\mathbb{Q}|\mathbb{P})&=\frac{q}{2}\mathbb{E}_\mathbb{P}\left[\int_t^T(D_{t, s}^{\mathbb{Q},\mathbb{P}})^q\big(|\lambda_s|^2+|\alpha^\mathbb{Q}_s|^2\big)ds~\Big|~\mathcal{F}_{t}\right],\label{hqt2}
		\end{align}
		where $\alpha^\mathbb{Q}$ is determined by \eqref{eq:dt}.
		
		\item[(ii)] In particular, we have that	
		\begin{equation}\label{eq:2}
		H_q(\mathbb{Q}|\mathbb{P})=\frac{q}{2}\mathbb{E}_\mathbb{P}\left[\int_0^T(D_s^{\mathbb{Q},\mathbb{P}})^q\big(|\lambda_s|^2+|\alpha^\mathbb{Q}_s|^2\big)ds\right].
		\end{equation}	
		Besides, we get
		\begin{equation}\label{eq:4}
		H_q(\mathbb{Q}^{min}|\mathbb{P})=\frac{q}{2}\mathbb{E}_\mathbb{P}\left[\int_0^T(D_s^{\mathbb{Q}^{min},\mathbb{P}})^q |\lambda_s|^2ds\right].
		\end{equation}	
	\end{itemize}
\end{theorem}

\begin{corollary}\label{qqmin}
	Let $q>0$ and $q\neq 1$. For any $\mathbb{Q}\in \mathcal{M}$, if $H_q(\mathbb{Q}|\mathbb{Q}^{min})<+\infty$, then,  for each $t\in[0,T]$, we have
	\begin{equation}\label{eq:3}
	H_{q,t}(\mathbb{Q}|\mathbb{Q}^{min})=	\frac{q}{2}\mathbb{E}_{\mathbb{Q}^{min}}\left[\int_t^T(D_{t, s}^{\mathbb{Q},\mathbb{Q}^{min}})^q\cdot |\alpha^\mathbb{Q}_s|^2 ds~\big|~\mathcal{F}_{t}\right],
	\end{equation}	and in particular, 	
	\begin{equation}\label{eq:qmin}	
	H_q(\mathbb{Q}|\mathbb{Q}^{min})=\frac{q}{2}\mathbb{E}_{\mathbb{Q}^{min}}\left[\int_0^T(D_s^{\mathbb{Q},\mathbb{Q}^{min}})^q\cdot \big|\alpha^\mathbb{Q}_s|^2ds\right],
	\end{equation}	
	where $\alpha^\mathbb{Q}$ is determined by \eqref{eq:dt}.	
\end{corollary}
\begin{remark}
	The readers can refer to  \cite{CS05} and  \cite{Skidas03} for the integral representation results of the classical relative entropy. When $q=1$, our representation results are exactly the corresponding formulations of classical relative entropy. 	
\end{remark}

\section{Main results}\label{sec3}

Before introducing our main results, let's introduce some notations.  Suppose $q>0$, $q\neq 1$, and the aversion coefficient $\gamma>0$.
We denote by $\mathcal{L}_{q}^{\gamma}(\mathcal{F}_T)$ for the set of all random variables on $(\Omega, \mathcal{F}_{T})$ such that $\exp_q(-\gamma \xi)\in L^{1}(\mathbb{P})$ and $\mathbb{E}_\mathbb{P}[\exp_q(-\gamma \xi)]\in \textrm{Dom}(\ln_{q})$. Define  $$\mathcal{L}_{q}^{\gamma}(\mathcal{F}_T, b):=\{  \xi\in \mathcal{L}_{q}^{\gamma}(\mathcal{F}_T)~|~ \exists ~ m_{1}, m_{2}\in \mathbf{R}, ~s. t.,~0<m_{1}\leq \exp_{q}(-\gamma \xi)\leq m_{2},~ \mathbb{P-}a.s.~\}.$$

Let $(\theta_t)_{t\in[0,T]}$ be a $\mathbf{R}^n$-valued locally square integrable predictable process on  $(\Omega, \mathcal{F}_T, \mathbb{P})$.  For each $t\in [0,T]$, define  $M_t:=\int_0^t\theta_s\cdot dW^{\perp}_s$. If the process $M$ is a BMO($\mathbb{P}$) martingale, then the stochastic exponential $\mathcal{E}(M)$ is a uniformly integrable $\mathbb{P}$-martingale by Theorem 2.3 in  \cite{K94}. Then we can define an equivalent martingale measure $\mathbb{Q}^\theta$ in $\mathcal{M}$ as follows:
\begin{align*}
\frac{~d\mathbb{Q}^{\theta}}{d\mathbb{P}}:&=\mathcal{E}(-\lambda\cdot W+\theta\cdot W^{\perp})_T\\
&=\exp\left (-\int_0^T\lambda_s\cdot dW_s+\int_0^T\theta_s\cdot dW^{\perp}_s-\frac{1}{2}\int_0^T(|\lambda_s|^2+|\theta_s|^2)ds\right).
\end{align*}

Denote $\Theta$ by the set of all $\mathbf{R}^n$-valued predictable processes $(\theta_{t})_{t\in[0.T]}$ such that $\int_0^\cdot\theta_s\cdot dW^{\perp}_s$ is a BMO($\mathbb{P}$) martingale and the following backward equation
\begin{align}\label{ty}
Y^{\theta}_{t}=\mathbb{E}_{\mathbb{Q}^{\theta}}\left[~\xi +\int_{t}^{T}\frac{\mu(Y^{\theta}_{s})}{2\gamma}|\theta_{s}|^{2}ds \Big| \mathcal{F}_{t} \right], ~~t\in[0,T], ~~\xi\in \mathcal{L}^{\gamma}_{q}(\mathcal{F}_T, b),
\end{align}
has a unique bounded solution $Y^{\theta}$ with $\mathbb{E}_{\mathbb{Q}^{\theta}}\left[\int_{0}^{T}|\mu(Y^{\theta}_{s})|\cdot |\theta_{s}|^{2}ds\right]<+\infty$, where $$\mu(x):=\frac{1-(1-q)\gamma x}{q}=\frac{1}{q}\big(\exp_q(-\gamma x)\big)^{1-q}. $$

The main result of the paper is the connection among the following three problems. 

\begin{itemize}
\item {\bf{Problem 1:}} For any contingent claim $\xi\in\mathcal{L}_{q}^{\gamma}(\mathcal{F}_T, b) $, pricing principle is defined by  
\begin{align}\label{motivation1}
F_t(\xi):= \mathop{\essinf}_{\mathbb{Q}\in \mathcal{M}_f} \left( \mathbb{E}_{\mathbb{Q}^{min}}\Big[(D_{t,T}^{\mathbb{Q},\mathbb{Q}^{min}})^q~\xi~\big|~\mathcal{F}_t\Big]+ \frac{1}{ \gamma} H_{q,t}(\mathbb{Q}|\mathbb{Q}^{min})\right), ~~~t\in[0,T],
\end{align}where $$\mathcal{M}_f:=\{\mathbb{Q}\in \mathcal{M}~|~ H_q(\mathbb{Q}|\mathbb{Q}^{min})<+\infty\}.$$
\item {\bf{Problem 2:}}   For any contingent claim $\xi\in\mathcal{L}_{q}^{\gamma}(\mathcal{F}_T, b) $, 
\begin{equation}\label{dualrep}
	\tilde{F}_t(\xi):= \mathop{\essinf}_{\theta\in \Theta} Y^{\theta}_{t}, ~~~t\in[0,T],
	\end{equation}where $Y^{\theta}$ is the solution of  \eqref{ty}.
\item {\bf{Problem 3:}}   Consider the following BSDE: for each  $\xi\in \mathcal{L}_{q}^{\gamma}(\mathcal{F}_T, b)$,
\begin{align}\label{bsde}
\left\{\begin{array}{lll}
d Y_{s}&=\left(\lambda_s\cdot  Z_s+\frac{\gamma }{2}\cdot\frac{ |Z_{s}^{\perp}|^{2}}{\mu( Y_{s})}\right)ds+Z_{s}\cdot dW_{s}+Z_s^{\perp}\cdot dW_s^{\perp},~~~s\in[0,T], \\   Y_{T}&=\xi.
\end{array}
\right.
\end{align}
\end{itemize}

\begin{theorem}\label{main}
	Suppose $\gamma>0$, $q>0$ and $q\neq 1$. For each  $\xi\in \mathcal{L}^{\gamma}_{q}(\mathcal{F}_T, b)$,  we have
\begin{align}\label{mainresult}
F_t(\xi)=\tilde{F}_t(\xi)=Y_t(\xi),   ~~~\forall t\in[0,T],
\end{align}where $Y_{\cdot}(\xi)$ is the solution of BSDE \eqref{bsde}.  Moreover, the essential infima in \eqref{motivation1} and \eqref{dualrep} can be achieved.
\end{theorem}

Our objective is to provide a pricing principle for the non-attainable contingent claim in incomplete market.   Problem 1 is  the starting point and motivation for our study  in the introduction.   This pricing rule is fascinating and attractive. The buyer evaluates the contingent claim $\xi$ under the ``distorted Radon-Nikodym derivative'' and adjustment by conditional Tsallis relative entropy over a family of equivalent martingale measures.
	
Specifically, the influence of distortion parameter can be understood in the following ways. When $0<q<1$, it enhances the small values of Radon-Nikodym derivative and reduces the larger values. However, when $q$ is greater than one, the opposite is true. Hence, we can interpret $q$ as a measure of agent's pessimism.

 To solve the Problem 1, we  introduce the Problem 2 and Problem 3.   Problem 2 is a dual formulation, where the value $Y^{\theta}$ has a recursive representation.  The buyer assesses the contingent claim  $\xi$ under a martingale measure $\mathbb{Q}^{\theta}$, corrects it by the backward equation  \eqref{ty}, and finally takes the worst case under a family of equivalent martingale  measures $\mathcal{M}^{\Theta}$,  which is defined by 
$$\mathcal{M}^{\Theta}=\{\mathbb{Q}^{\theta}\in \mathcal{M}~|~ \theta\in \Theta \}.$$
Then the pricing problem  \eqref{dualrep} becomes an optimization problem of BSDEs.

The dynamic formulation in BSDE \eqref{bsde} of Problem 3 is important, because it is not clear that $F_{\cdot}(\xi)$ defined in \eqref{motivation1} is time consistent, due to the time subscript $t$ in $D_{t,T}^{\mathbb{Q},\mathbb{Q}^{min}}$ and the distortion parameter $q$.  Similarly, we cannot easily obtain that $\tilde{F}_{\cdot}(\xi)$ is dynamically consistent because it is also an optimization problem for a series of recursions.  The optimal value process is described as the solution of the BSDE \eqref{bsde}.

The kind of BSDEs with generators $f(y)|z|^2$  are firstly investigated by  \cite{BEO17}, and  subsequently studied by  \cite{BT18},  \cite{B20} and  \cite{ZZF21}.  The existence of solution to BSDE \eqref{bsde} can be obtained by the existing literature,  while the uniqueness part  of  BSDE \eqref{bsde}  is not available in the current literature.   Since the generator of BSDE \eqref{bsde} is concave in $(y,z^\perp)$,  motivated by the $\theta$-method in  \cite{BH08} dealing with the convex generator, we can derive the uniqueness result by some subtle transformations. 
We  give it  in the following proposition.

\begin{proposition}\label{bsde:eq2} Suppose $\gamma>0$, $q>0$ and $q\neq 1$. For any  $\xi\in \mathcal{L}_{q}^{\gamma}(\mathcal{F}_T, b)$, then BSDE \eqref{bsde} admits a unique solution $(Y, \overline{Z})=(Y, Z, Z^{\perp})$ in which $\int_{0}^{\cdot}\overline{Z}_{s}\cdot d\overline{W}_s$ is a BMO($\mathbb{P}$) martingale, and $Y$ is continuous and bounded, specifically, for each $t\in[0,T]$, $Y_t\in \mathcal{L}_{q}^{\gamma}(\mathcal{F}_t, b)$. 	
\end{proposition}

On the one hand,  by Proposition \ref{bsde:eq2} and Theorem \ref{main},  for each $t\in[0,T]$,  the pricing principle is a mapping  that   
$F_t: \mathcal{L}_{q}^{\gamma}(\mathcal{F}_T, b)\rightarrow \mathcal{L}_{q}^{\gamma}(\mathcal{F}_t, b)$ by
\begin{equation*}
F_t(\xi)=Y_t(\xi)=Y_t,  ~~~~\xi\in \mathcal{L}_{q}^{\gamma}(\mathcal{F}_T,b),
\end{equation*}where $Y_{\cdot}$ is the solution for BSDE \eqref{bsde}. In particular, $F(\xi):=F_0(\xi)=Y_0(\xi)=Y_0$. We often omit $\xi$ when there is no ambiguity.   BSDE formulation  \eqref{bsde} and  the result in \eqref{mainresult}  indicate that $F_{\cdot}(\xi)$ is  a \textit{time consistent} pricing rule.

On the other hand,  for any contingent claim $\xi\in \mathcal{L}_{q}^{\gamma}(\mathcal{F}_T, b)$, the solution of BSDE \eqref{bsde} satisfies the fact that  $\int_{0}^{\cdot}\overline{Z}_{s}\cdot d\overline{W}_s$ is a BMO($\mathbb{P}$) martingale and $Y_t\in \mathcal{L}_{q}^{\gamma}(\mathcal{F}_t, b)$, $\forall t\in[0,T]$.  These results imply that $\mu(Y)$ is bounded (strict greater than a positive constant) and $\int_0^\cdot Z^{\perp}_s\cdot dW^\perp_s$ is a BMO$(\mathbb{P})$ martingale, then  $Y_\cdot$ is a martingale under the equivalent martingale measure $\mathbb{Q}^{\xi}$ in $\mathcal{M}$, i.e., 
$$Y_t(\xi)=Y_t=\mathbb{E}_{\mathbb{Q}^{\xi}}[~\xi~|~\mathcal{F}_t],~~0\leq t\leq T,$$
where $$\frac{d\mathbb{Q}^{\xi}}{d\mathbb{P}}:=D_T^{\mathbb{Q^{\xi},\mathbb{P}}}=\mathcal{E}(-\lambda\cdot W-\frac{\gamma Z^{\perp} }{2\mu(Y)}\cdot W^\perp)_T.$$
Hence, the pricing principle $F$ for the contingent claim $\xi$ is an \textit{arbitrage-free} pricing rule.   

This pricing principle, connected with the above BSDEs,  is very interesting and exciting.  The pricing mechanism is nonlinear yielding the price in terms of a conditional nonlinear expectation. Since BSDE's generator  involves  $y$, the pricing principle will not have cash additivity or translation invariance property generally. Translation invariance has always been controversial (for example,  \cite{ER09}, \cite{HWWX21}), and our  pricing principle will shed some lights on this direction.

We end this section with some useful remarks.

\begin{remark}
	The pricing principle \eqref{motivation1} is compatible with risk neutral pricing principle for the attainable contingent claim. It is also related to the conditional certainty equivalent of some utility function when the contingent claim is completely unhedged. 
	
	\begin{itemize}
		\item If the contingent claim is \textit{attainable}, i.e., $\xi\in \mathcal{L}_{q}^{\gamma}(\mathcal{F}^{W}_T,b)$,  then $Z^{\perp}=0$, and
		$$F_t(\xi)=\mathbb{E}_{\mathbb{Q}^{min}}[~\xi~|\mathcal{F}_t^{W}],~~~ t\in[0,T].$$Then this pricing functional is compatible with risk neutral pricing principle.

		\item If the contingent claim is \textit{completely unhedged}, i.e., $\xi\in \mathcal{L}_{q}^{\gamma}(\mathcal{F}^{W^{\perp}}_T,b)$, then $Z=0$, and
		\begin{align}\label{xiangdeng}
		F_t(\xi)&=-\frac{1}{\gamma} \ln_q\mathbb{E}_{\mathbb{Q}^{min}}[\exp_q(-\gamma\xi)~|\mathcal{F}_t^{W^{\perp}}]\\
		&=U^{-1}(\mathbb{E}_{\mathbb{Q}^{min}}[~U(\xi)~|\mathcal{F}_t^{W^{\perp}}]),\quad\quad~~~ t\in[0,T].\nonumber
		\end{align}where $U$ is the $q$-exponential utility, i.e. $U(x):=1-\exp_{q}(-\gamma x)$.
		In this situation, $F_t(\xi)$ equals to the conditional certainty equivalent under $q$-exponential utility.  This result will be expanded further in subsection \ref{sec4.3}. 
\item   For the \textit{general} contingent claim $\xi\in\mathcal{L}_{q}^{\gamma}(\mathcal{F}_T, b)$,  we will prove that
	$$
	U^{-1}(\mathbb{E}_{\mathbb{Q}^{min}}[~U(\xi)~|\mathcal{F}_t])	\leq F_t(\xi)\leq \mathbb{E}_{\mathbb{Q}^{min}}[\xi\big|\mathcal{F}_t],  ~~~ t\in[0,T],
	$$where $U$ is the $q$-exponential utility, see Proposition \ref{jixian23}.
	\end{itemize}
\end{remark}

\begin{remark}	
	When $q=1$ and the contingent claim is  \textit{completely unhedged}, i.e.,  $\xi\in L^{\infty}(\mathcal{F}^{W^{\perp}}_T)$, by Proposition 6.4 in  \cite{BK09}, we have known that this pricing functional is the conditional certainty equivalent of exponential utility under the minimal martingale measure, i.e.,
	\begin{align}
	F_t(\xi)=-\frac{1}{\gamma} \ln\mathbb{E}_{\mathbb{Q}^{min}}[\exp(-\gamma\xi)~|\mathcal{F}_t^{W^{\perp}}], ~~~ t\in[0,T].
	\end{align}
\end{remark}

\begin{remark} For each contingent claim  $\xi\in \mathcal{L}_{q}^{\gamma}(\mathcal{F}_T, b)$,   the essential infimum in \eqref{motivation1} can be achieved at the equivalent martingale measure $\mathbb{Q}^*\in \mathcal{M}_f$, i.e., 
$$\frac{d\mathbb{Q}^*}{d\mathbb{P}}=\mathcal{E}(-\lambda\cdot W+\alpha^{\mathbb{Q}^*}\cdot {W}^{\perp})_T,$$where 
$\alpha^{\mathbb{Q}^*}=-\frac{\gamma Z^{\perp}}{q\mu(Y)}.$  While the essential infimum in \eqref{dualrep}  can be obtained by 
$$\theta^{*}=-\frac{\gamma Z^\perp}{\mu(Y)}.$$
The readers can find these results in the proofs process of Theorem \ref{main}, see  Lemma \ref{dual} and Lemma \ref{dual-tsallis} in  Appendix \ref{seca2}.
\end{remark}

\begin{remark} 
For each $t\in[0,T]$, the pricing principle $F_t(\xi)$ is \textit{the price to buy} the contingent claim $\xi$ for DM.  Similarly, we can also define \textit{the price to sell} the claim $\xi$ in $\mathcal{L}_{q}^{-\gamma}(\mathcal{F}_T, b)$ by 
\begin{align}\label{motivation2}
-F_t(-\xi):= \mathop{\esssup}_{\mathbb{Q}\in \mathcal{M}_f} \left( \mathbb{E}_{\mathbb{Q}^{min}}\Big[(D_{t,T}^{\mathbb{Q},\mathbb{Q}^{min}})^q~\xi~\big|~\mathcal{F}_t\Big]- \frac{1}{ \gamma} H_{q,t}(\mathbb{Q}|\mathbb{Q}^{min})\right).
\end{align}In this paper, we focus on the buyer's price $F$.
\end{remark}

\section{Some properties of  the pricing principle}\label{sec4}

The goal of this section is to study the properties for the pricing principle $F$. In subsection \ref{sec4.1}, we introduce the basic properties of the pricing rule.   We discuss its relationship with  conditional certainty equivalent under $q$-exponential utility in subsection \ref{sec4.3}. Subsection \ref{sec4.2} considers  the asymptotic behavior of the pricing principle when the aversion coefficient goes to zero or infinity.

\subsection{Basic properties}\label{sec4.1}

We claim that  the pricing principle satisfies
normalization, monotonicity, concavity, time consistency, and cash subadditivity or superadditivity and so on.

\begin{proposition}\label{property1} Let $\gamma>0$, $q>0$ and $q\neq 1$. For each $t\in[0,T]$, the pricing principle $F_t: \mathcal{L}_{q}^{\gamma}(\mathcal{F}_T, b)\rightarrow \mathcal{L}_{q}^{\gamma}(\mathcal{F}_t, b)$, defined in \eqref{motivation1}, satisfies the following properties:
	\begin{itemize}
		\item[(i)] Normalization: $F_t(0)=0$.
		\item[(ii)] Monotonicity: For $\xi, \eta\in \mathcal{L}_{q}^{\gamma}(\mathcal{F}_T, b)$ with $\xi\geq \eta$, then $F_t(\xi)\geq F_t(\eta)$. 
		\item[(iii)] Concavity: If $\kappa \xi+(1-\kappa)\eta  \in \mathcal{L}_{q}^{\gamma}(\mathcal{F}_T, b)$ for all $\kappa\in[0,1]$, then $$F_t(\kappa \xi+(1-\kappa)\eta)\geq \kappa F_t(\xi)+(1-\kappa) F_t(\eta).$$
		\item[(iv)] Quasi-concavity: If $\kappa \xi+(1-\kappa)\eta  \in \mathcal{L}_{q}^{\gamma}(\mathcal{F}_T, b)$ for all $\kappa\in[0,1]$, then $$F_t(\kappa \xi+(1-\kappa)\eta)\geq   \min\{ F_t(\xi), F_t(\eta)\}.$$
		\item[(v)] Scaling property:  Let $\kappa \xi, \xi\in \mathcal{L}_{q}^{\gamma}(\mathcal{F}_T, b)$.  Then $F_t(\kappa \xi)\geq \kappa F_t(\xi)$ if $\kappa\in[0,1]$; $F_t(\kappa \xi)\leq \kappa F_t(\xi)$ if  $\kappa\geq 1$.
		\item[(vi)] Time consistency: For $\xi\in \mathcal{L}_{q}^{\gamma}(\mathcal{F}_T, b)$,  then $$F_s(\xi)=F_s(F_t(\xi)),~~~~0\leq s \leq t \leq T.$$
	\end{itemize}	
\end{proposition}
\textbf{Proof.}
	(i)-(iii) can be easily verified by the definition in \eqref{motivation1}.  (iv) and (v) follow directly from (iii) and (i).  (vi) uses the results in Theorem \ref{main}. \hfill$\Box$

\begin{proposition}\label{property2} Let $\gamma>0$, $q>0$ and $q\neq 1$. For each $t\in[0,T]$, given the pricing principle $F_t: \mathcal{L}_{q}^{\gamma}(\mathcal{F}_T, b)\rightarrow \mathcal{L}_{q}^{\gamma}(\mathcal{F}_t, b)$, defined in \eqref{motivation1}. Let $c$ be a constant,
	and  $\xi, \xi+c\in \mathcal{L}_{q}^{\gamma}(\mathcal{F}_T, b)$. Then we have that	
	\begin{itemize}
		\item[(i)] Cash superadditivity:  If $0<q<1$, $c\leq0$ or $q>1$, $c\geq0$,
		then
		$$F_t(\xi+c)\geq F_t(\xi)+c.$$
		\item[(ii)] Cash subadditivity:  If $0<q<1$, $c\geq0$ or $q>1$, $c\leq0$, then
		$$F_t(\xi+c)\leq F_t(\xi)+c.$$
	\end{itemize}			
\end{proposition}
\textbf{Proof.}  Case (i): If $0<q<1$, $c\leq0$,  by the definition of  the pricing principle, 
	it implies 
	\begin{align*}
	F_t(\xi+c)&= \mathop{\essinf}_{\mathbb{Q}\in \mathcal{M}_f} \left(\mathbb{E}_{\mathbb{Q}^{min}}\left[(D_{t,T}^{\mathbb{Q},\mathbb{Q}^{min}})^q~(\xi+c)~\big|~\mathcal{F}_t\right]++\frac{1}{\gamma}H_{q,t}(\mathbb{Q}~|~\mathbb{Q}^{min}) \right)\\
	&\geq F_t(\xi)+\mathop{\essinf}_{\mathbb{Q}\in \mathcal{M}_f}  \mathbb{E}_{\mathbb{Q}^{min}}\left[c\cdot(D_{t,T}^{\mathbb{Q},\mathbb{Q}^{min}})^q~\big|~\mathcal{F}_t\right]\\
	&\geq F_t(\xi)+ \mathop{\essinf}_{\mathbb{Q}\in \mathcal{M}_f}  c\cdot\left(\mathbb{E}_{\mathbb{Q}^{min}}\left[D_{t,T}^{\mathbb{Q},\mathbb{Q}^{min}}~\big|~\mathcal{F}_t\right]\right)^q\\
	&=F_t(\xi)+ c,
	\end{align*}
	where we used the Jensen's inequality. When $q>1$ and $c\geq0$, the proof is similar.

	Case (ii):  If $0<q<1$, $c\geq0$, we get
	\begin{align*}
	F_t(\xi+c)&\leq F_t(\xi)+\mathop{\esssup}_{\mathbb{Q}\in \mathcal{M}_f}  \mathbb{E}_{\mathbb{Q}^{min}}\left[c\cdot(D_{t,T}^{\mathbb{Q},\mathbb{Q}^{min}})^q~\big|~\mathcal{F}_t\right]\\
	&\leq F_t(\xi)+ \mathop{\esssup}_{\mathbb{Q}\in \mathcal{M}_f}  c\cdot\left(\mathbb{E}_{\mathbb{Q}^{min}}\left[D_{t,T}^{\mathbb{Q},\mathbb{Q}^{min}}~\big|~\mathcal{F}_t\right]\right)^q\\
	&=F_t(\xi)+ c.
	\end{align*}When $q>1$ and $c\leq0$, the proof is similar.   \hfill$\Box$

\begin{remark}  \cite{DPG10} study dynamic convex risk measure or utility process, satisfying cash additivity,  and they find that it is related to a kind of BSDEs with generators are independent from $y$.  Cash or constant additivity has always been a controversial property for contingent claim pricing in academia. In order to model stochastic or ambiguous interest rates, \cite{ER09} present a kind of the cash subadditive risk measures. Our pricing principle $F$ is induced by Tsallis relative entropy theory, and it turns out to be related to a specific BSDE \eqref{bsde} involving $y$.  From the results of  Propositions \ref{property1} and  \ref{property2}, pricing principle $F$ has very good properties and is a suitable choice for contingent claim pricing.
	
\end{remark}

\subsection{Conditional certainty equivalent under $q$-exponential utility}\label{sec4.3}

Let $\gamma>0$, $q>0$ and $q\neq 1$. Consider the following function:
$$U(x):=1-\exp_{q}(-\gamma  x),~x\in \textrm{Dom}(U),$$which is called  $q$-exponential utility.
Obviously, $U$ is a strictly increasing and concave continuous function defined on $\textrm{Dom}(U)$.

For each $t\in[0,T]$, we consider the conditional certainty equivalent of the law of $\xi$ under minimal martingale measure $\mathbb{Q}^{min}$ by setting
$$
CE_{q}(\xi|\mathcal{F}_t):=U^{-1}(\mathbb{E}_{\mathbb{Q}^{min}}[U(\xi)|\mathcal{F}_t]), ~~~~\forall\xi\in\mathcal{L}_{q}^{\gamma}(\mathcal{F}_T,b).$$
Directly calculation results in the following
\begin{equation}\label{xiaoyongdengjia1}
CE_{q}(\xi|\mathcal{F}_t)=-\frac{1}{\gamma}\ln_{q}\mathbb{E}_{\mathbb{Q}^{min}}[\exp_q(-\gamma \xi)|\mathcal{F}_t], ~~~\forall\xi\in\mathcal{L}_{q}^{\gamma}(\mathcal{F}_T,b).
\end{equation}When $t=0$, we denote $CE_{q}(\xi|\mathcal{F}_0)$ by $CE_{q}(\xi)$.

By Theorem 4.2 in  \cite{MT21} or Theorem 3.2 in  \cite{ZZF21}, we can easily obtain that for each contingent claim $\xi\in \mathcal{L}_{q}^{\gamma}(\mathcal{F}_T, b)$, then
\begin{equation}\label{xiaoyongdengjia2}
CE_{q}(\xi|\mathcal{F}_t)=y_t,  ~~\forall t\in[0,T],
\end{equation}
where $y$ is the unique solution to the following BSDE
\begin{align}\label{bsde-zz}
\left\{\begin{array}{lll}
d y_{s}&=\left(\lambda_s\cdot  z_s+\frac{\gamma }{2}\cdot\frac{ |z_{s}|^{2}+|z_{s}^{\perp}|^{2}}{\mu( y_{s})}\right)ds+z_{s}\cdot dW_{s}+z_s^{\perp}\cdot dW_s^{\perp},~~~s\in[0,T], \\   y_{T}&=\xi.
\end{array}
\right.
\end{align}
Comparing with BSDE \eqref{bsde}, BSDE \eqref{bsde-zz} is easier to solve and it has an explicit solution characterized by conditional certainty equivalent of $q$-exponential utility.
BSDE \eqref{bsde-zz} admits a unique solution $(y,\overline{z})=(y, z,z^{\perp})$ in which $y$ is continuous and bounded, and
$\int_{0}^{\cdot}\overline{z}_{s}\cdot d\overline{W}_s^{min}$ is a BMO($\mathbb{Q}^{min}$) martingale, where $\overline{W}^{min}=(W^{-\lambda}, W^{\perp})$ is a Brownian motion under $\mathbb{Q}^{min}$, and $W_{\cdot}^{-\lambda}=W_{\cdot}+\int_0^{\cdot}\lambda_sds$.

\begin{remark}
	 \cite{MT21} prove that the solution of BSDE \eqref{bsde-zz} is conditional certainty equivalent of $q$-exponential utility, however,  they do not investigate its representation and not explore its  potential applications in pricing theory.
	
\end{remark}

In general, conditional certainty equivalent of $q$-exponential utility is not compatible with the risk neutral pricing for attainable claims.
In fact,
$$CE_{q}(\xi|\mathcal{F}_{t})=\mathbb{E}_{\tilde{\mathbb{Q}}^{\xi}}[~\xi~|~\mathcal{F}_t],~~~\forall t\in[0,T],$$
where $$\frac{d\tilde{\mathbb{Q}}^{\xi}}{d\mathbb{P}}:=D_T^{\tilde{\mathbb{Q}}^{\xi},\mathbb{P}}=\mathcal{E}(-\lambda\cdot W-\frac{\gamma z }{2\mu(y)}\cdot W-\frac{\gamma z^{\perp} }{2\mu(y)}\cdot W^{\perp})_T.$$
The equivalent measure $\tilde{\mathbb{Q}}^{\xi}$ is \textit{not} an equivalent martingale measure if $z\neq 0$.   For example,  for some $\xi$ in $\mathcal{L}_{q}^{\gamma}(\mathcal{F}^{W}_T,b)$.

When the contingent claim is completely unhedged, i.e., $\xi\in \mathcal{L}_{q}^{\gamma}(\mathcal{F}^{W^{\perp}}_T,b)$, then $Z=z=0$, where $Z$ and $z$ are part of solutions to BSDEs \eqref{bsde} and \eqref{bsde-zz} respectively. In this case, we get \eqref{xiangdeng} holds, i.e.,
\begin{align}
F_t(\xi)=CE_{q}(\xi|\mathcal{F}_t), ~~\forall t\in[0,T].
\end{align}	
However, they are not equal to each other generally.

Define
$$\mathcal{Q}_f:=\{\mathbb{Q}\in \mathcal{Q}~|~ H_q(\mathbb{Q}|\mathbb{Q}^{min})<+\infty\}.$$The following Proposition \ref{dual-tsallis-z}  further  reveals the differences and connections between pricing principles $F_{t}(\xi)$ and $CE_{q}(\xi|\mathcal{F}_{t})$.

\begin{proposition}\label{dual-tsallis-z}
	Suppose $\gamma>0$, $q>0$ and $q\neq 1$. For any $\xi \in\mathcal{L}^{\gamma}_{q}(\mathcal{F}_T, b)$ and  $t\in[0,T]$, then we have
	\begin{align}
	CE_{q}(\xi|\mathcal{F}_t)
	&=\mathop{\essinf}_{\mathbb{Q}\in \mathcal{Q}_f} \mathbb{E}_{\mathbb{Q}^{min}}\left[(D_{t,T}^{\mathbb{Q},\mathbb{Q}^{min}})^q~\xi~+\frac{q}{2 \gamma}\int_{t}^{T}(D_{t,s}^{\mathbb{Q},\mathbb{Q}^{min}})^q(|\beta_s^{\mathbb{Q}}|^2+|\alpha^{\mathbb{Q}}_s|^2)ds~\big|~\mathcal{F}_t\right]\label{dualrep11z}\\
&=\mathop{\essinf}_{\mathbb{Q}\in \mathcal{Q}_f} \left(\mathbb{E}_{\mathbb{Q}^{min}}\left[(D_{t,T}^{\mathbb{Q},\mathbb{Q}^{min}})^q~\xi~\big|~\mathcal{F}_t\right]+\frac{1}{ \gamma}H_{q,t}(\mathbb{Q}|\mathbb{Q}^{min})\right), \label{ceqt}
	\end{align}where $\beta^{\mathbb{Q}}$ and $\alpha^{\mathbb{Q}}$ are determined by $D^{\mathbb{Q},\mathbb{Q}^{min}}=\mathcal{E}(\beta^{\mathbb{Q}}\cdot W^{-\lambda}+\alpha^\mathbb{Q}\cdot W^\perp)$. Moreover, the essential infimum  can be attained.
	In particular,
	\begin{align}
	CE_{q}(\xi)&= \mathop{\inf}_{\mathbb{Q}\in \mathcal{Q}_f} \left( \mathbb{E}_{\mathbb{Q}^{min}}\big[(D_{T}^{\mathbb{Q},\mathbb{Q}^{min}})^q~\xi~\big]+\frac{1}{ \gamma} H_q(\mathbb{Q}|\mathbb{Q}^{min}) \label{dualrep22z}\right).
	\end{align}
\end{proposition}

By comparing  Proposition \ref{dual-tsallis-z} with \eqref{motivation1}, we find that conditional certainty equivalent pricing is a robust representation over a family of equivalent measures in $\mathcal{Q}_f$, while $F$ is a robust representation over a family of equivalent martingale measures in $\mathcal{M}_f$. Combining \eqref{motivation1} and \eqref{ceqt} together, we can finally obtain the following results. 

\begin{proposition}\label{jixian23} Let $\gamma>0$, $q>0$ and $q\neq 1$. For each $t\in[0,T]$ and   $\xi\in\mathcal{L}_{q}^{\gamma}(\mathcal{F}_T, b)$,  then it implies 
	\begin{equation}\label{jie2}
	CE_{q}(\xi|\mathcal{F}_t)	\leq F_t(\xi)\leq \mathbb{E}_{\mathbb{Q}^{min}}[\xi\big|\mathcal{F}_t].
	\end{equation}
\end{proposition}

Proposition \ref{jixian23} indicates that  our pricing rule $F$ is between minimal martingale measure pricing and conditional certainty equivalent pricing under  $q$-exponential utility function. 

\begin{remark}
	The sell price of the contingent claim $\xi$ can be studied similarly. In fact, for each $t\in[0,T]$ and   $\xi \in\mathcal{L}_{q}^{\gamma}(\mathcal{F}_T, b)\cap\mathcal{L}_{q}^{-\gamma}(\mathcal{F}_T, b)$,  then
	\begin{equation}\label{jie23}
	CE_{q}(\xi|\mathcal{F}_t)	\leq F_t(\xi)\leq \mathbb{E}_{\mathbb{Q}^{min}}[\xi\big|\mathcal{F}_t]\leq -F_t(-\xi)\leq -CE_{q}(-\xi|\mathcal{F}_t).
	\end{equation}
	Specifically, $[F_t(\xi), -F_t(-\xi)]$ is the arbitrage-free pricing interval for  contingent claim $\xi$.
\end{remark}

\subsection{The effects of  ambiguity aversion coefficient }\label{sec4.2}

In order to highlight the influence of the aversion parameter $\gamma$, we denote $F_t(\xi)$ by
$F_t(\xi,\gamma)$ for any $\xi \in \mathcal{L}_{q}^{\gamma}(\mathcal{F}_T, b)$ in this subsection.

\begin{proposition}\label{property3} Let $\gamma>0$,  $q>0$ and $q\neq 1$. For each $t\in[0,T]$, the pricing principle $F_t(\xi, \gamma)$ has the following properties.
	\begin{itemize}
		\item[(i)] It is decreasing in  $\gamma$, i.e., if $\gamma\leq \gamma'$ and $\xi\in\mathcal{L}_{q}^{\gamma}(\mathcal{F}_T, b)\cap \mathcal{L}_{q}^{\gamma'}(\mathcal{F}_T, b)$,
		then we have
		$$F_t(\xi,\gamma')\leq F_t(\xi,\gamma).$$
		\item[(ii)] For any $\kappa>0$, if $\kappa\xi\in \mathcal{L}_{q}^{\gamma}(\mathcal{F}_T, b)$, then
		$$F_t(\kappa\xi,\gamma)= \kappa F_t(\xi,\kappa\gamma).$$
	\end{itemize}			
\end{proposition}
\textbf{Proof.}
	For any $\kappa>0$, if $\kappa\xi\in \mathcal{L}_{q}^{\gamma}(\mathcal{F}_T, b)$, then it implies that $\xi\in \mathcal{L}_{q}^{\kappa\gamma}(\mathcal{F}_T, b)$. Thus, the results are immediately consequences of the definition of the pricing principle $F$.    \hfill$\Box$


The readers can refer to  \cite{REK00} and  \cite{MS05}, and these authors also give the asymptotic results for large and small aversion coefficients of indifference pricing theory in exponential utility and relative entropy framework.  

The following proposition is the asymptotic results of our pricing principle.

\begin{proposition}\label{jixian} 
Let $q>0$ and $q\neq 1$. For each $t\in[0,T]$ and   $\xi\in\mathop{\bigcap}_{\gamma>0}\mathcal{L}_{q}^{\gamma}(\mathcal{F}_T, b)$,  then we have
\begin{equation}\label{gamma1}
\lim_{\gamma\rightarrow +\infty}F_t(\xi,\gamma)=\mathop{\essinf}_{\mathbb{Q}\in \mathcal{M}_f} \mathbb{E}_{\mathbb{Q}^{min}}\left[(D_{t,T}^{\mathbb{Q},\mathbb{Q}^{min}})^q~\xi~\big|~\mathcal{F}_t\right], \quad \mathbb{P}-a.s.\end{equation}
 and
\begin{equation}\label{gamma0}\lim_{\gamma\rightarrow 0}F_t(\xi,\gamma)=\mathbb{E}_{\mathbb{Q}^{min}}[\xi\big|\mathcal{F}_t],  \quad \mathbb{P}-a.s.
\end{equation}
\end{proposition}

Loosely speaking, the interpretation of Proposition \ref{jixian} is that, in the small risk
aversion limit, our pricing principle converges to minimal martingale measure pricing; in the large ambiguity aversion limit,   our pricing principle converges to the ``distorted Radon-Nikodym
derivative'' pricing over a family of equivalent martingale measures $\mathcal{M}_f$.  By Proposition \ref{jixian}, we can immediately draw the following corollary.

\begin{corollary}\label{jixian2} Let $\gamma>0$, $q>0$ and $q\neq 1$. For each $t\in[0,T]$ and   $\xi\in\mathcal{L}_{q}^{\gamma}(\mathcal{F}_T, b)$,  then
	\begin{equation}\label{jie1}
	\mathop{\essinf}_{\mathbb{Q}\in \mathcal{M}_f} \mathbb{E}_{\mathbb{Q}^{min}}\left[(D_{t,T}^{\mathbb{Q},\mathbb{Q}^{min}})^q~\xi~\big|~\mathcal{F}_t\right]	\leq F_t(\xi,\gamma)\leq \mathbb{E}_{\mathbb{Q}^{min}}[\xi\big|\mathcal{F}_t].
	\end{equation}
\end{corollary}

\section{Conclusions}\label{sec6}

The paper introduces a pricing principle for contingent claims in incomplete markets via Tsallis relative entropy theory. The agent evaluates the contingent claim under the ``distorted Radon-Nikodym
derivative'' and adjustment by Tsallis relative entropy over a family of equivalent martingale measures.   In order to investigate the pricing principle,  we introduce two equivalent problems of BSDEs.   
Fortunately,  similar to connections of quadratic BSDEs and relative entropy theory (exponential utility pricing theory),  BSDEs with generators $f(y)|z|^2$ type are found to be closely related to this pricing criterion.  

 The pricing principle  processes lots of elegant properties, such as monotonicity, time consistency, concavity and so on. It is also compatible with risk neutral pricing principle when contingent claim is attainable. Our pricing principle sheds some lights on  arbitrage-free pricing without translation invariance.   This  pricing criterion is a suitable choice because it is between the minimal martingale measure pricing and the conditional certainty equivalent pricing. We also consider  the asymptotic behavior of the pricing principle when the ambiguity aversion coefficient goes to zero or infinity.


\section{Appendix: The Proofs} \label{sec5}
In this appendix, we will give the proofs of the previous theorems and propositions.

\subsection{Proof of  Theorem \ref{shang1}}

\begin{lemma}\label{lemma1}
	
	Let $q>0$ and $q\neq 1$. For $\mathbb{Q}\ll \mathbb{P}$,  $D_\cdot^{\mathbb{Q},\mathbb{P}}=\mathbb{E}_\mathbb{P}[\frac{d\mathbb{Q}}{d\mathbb{P}}| \mathcal{F}_\cdot]$,  suppose that  $H_q(\mathbb{Q}|\mathbb{P})<+\infty$. Then  $(D^{\mathbb{Q},\mathbb{P}})^q\ln_qD^{\mathbb{Q},\mathbb{P}}$ is a $\mathbb{P}$-submartingale on $[0,T]$. Furthermore, $(D^{\mathbb{Q},\mathbb{P}})^q\ln_qD^{\mathbb{Q},\mathbb{P}}$ is a uniformly integrable $\mathbb{P}$-submartingale.
\end{lemma}
\textbf{Proof.}  Since the process $D^{\mathbb{Q},\mathbb{P}}$ is a martingale and $f(x)=x^{q}\ln_{q}(x)$,  $x\in\textrm{Dom}(\ln_{q})$, is convex and bounded
	from below, then Jensen's inequality yields, for $0\leq t\leq T$,
	$$\mathbb{E}_{\mathbb{P}}[f(D_T^{\mathbb{Q},\mathbb{P}})|\mathcal{F}_t]\geq f(\mathbb{E}_{\mathbb{P}}[D_T^{\mathbb{Q},\mathbb{P}}|\mathcal{F}_t])=f(D_t^{\mathbb{Q},\mathbb{P}}).$$Since $H_q(\mathbb{Q}|\mathbb{P})=\mathbb{E}_\mathbb{P}[(D_T^{\mathbb{Q},\mathbb{P}})^q\ln_qD_T^{\mathbb{Q},\mathbb{P}}]<+\infty$,  then $(D^{\mathbb{Q},\mathbb{P}})^q\ln_qD^{\mathbb{Q},\mathbb{P}}$ is a uniformly integrable $\mathbb{P}$-submartingale on $[0,T]$. \hfill $\Box$

\noindent {\bf{Proof of  Theorem \ref{shang1}.} }
Due to the fact that
	$$H_q(\mathbb{Q}|\mathbb{P})=\mathbb{E}_\mathbb{P}[(D_T^{\mathbb{Q},\mathbb{P}})^q\ln_qD_T^{\mathbb{Q},\mathbb{P}}]=\frac{1-\mathbb{E}_\mathbb{P}[(D_T^{\mathbb{Q},\mathbb{P}})^q]}{1-q}<+\infty,$$we have  $D_T^{\mathbb{Q},\mathbb{P}}\in L^{1\vee q}(\mathbb{P})$. 
	Let's prove \eqref{hqt1}.  Indeed,
	\begin{align*}
	H_{q,t}(\mathbb{Q}|\mathbb{P})=&\mathbb{E}_\mathbb{P}\Big[(D_{t,T}^{\mathbb{Q},\mathbb{P}})^q\ln_qD_{t, T}^{\mathbb{Q},\mathbb{P}}~\big|~\mathcal{F}_{t}\Big]\\
	=&\mathbb{E}_\mathbb{P}\left[\frac{D_{t,T}^{\mathbb{Q},\mathbb{P}}-(D_{t,T}^{\mathbb{Q},\mathbb{P}})^{q}}{1-q}~\Big|~\mathcal{F}_{t}\right]\\
	=&\frac{1}{(D_{t}^{\mathbb{Q},\mathbb{P}})^{q}}\mathbb{E}_\mathbb{P}\left[\frac{(D_{t}^{\mathbb{Q},\mathbb{P}})^{q}-(D_{T}^{\mathbb{Q},\mathbb{P}})^{q}}{1-q}~\Big|~\mathcal{F}_{t}\right]\\
	=&\frac{1}{(D_{t}^{\mathbb{Q},\mathbb{P}})^{q}}\mathbb{E}_\mathbb{P}\left[(D_{T}^{\mathbb{Q},\mathbb{P}})^q\ln_qD_{T}^{\mathbb{Q},\mathbb{P}}-(D_{t}^{\mathbb{Q},\mathbb{P}})^q\ln_qD_{t}^{\mathbb{Q},\mathbb{P}}~\Big|~\mathcal{F}_{t}\right],
	\end{align*}where the third equality is derived from $\mathbb{E}_\mathbb{P}[D_{t, T}^{\mathbb{Q},\mathbb{P}}~|~\mathcal{F}_{t}]=1$.

	Since
	$$dD_t^{\mathbb{Q},\mathbb{P}}=D_t^{\mathbb{Q},\mathbb{P}}(-\lambda_t\cdot dW_t+\alpha_t^\mathbb{Q}\cdot dW_t^\perp),~~ D_0^{\mathbb{Q},\mathbb{P}}=1,$$
	applying It\^{o}'s formula to $(D^{\mathbb{Q},\mathbb{P}})^q\ln_qD^{\mathbb{Q},\mathbb{P}}$, it yields that
	\begin{align*}
	d(D_t^{\mathbb{Q},\mathbb{P}})^q\ln_qD_t^{\mathbb{Q},\mathbb{P}}&=\frac{q}{2}(D_t^{\mathbb{Q},\mathbb{P}})^q\big(|\lambda_t|^2+|\alpha^\mathbb{Q}_t|^2\big)dt\\
	&~~~~+\frac{D_t^{\mathbb{Q},\mathbb{P}}-q(D_t^{\mathbb{Q},\mathbb{P}})^q}{1-q}(-\lambda_t\cdot dW_t+\alpha_t^\mathbb{Q}\cdot dW_t^\perp)\\
	&=\frac{q}{2}(D_t^{\mathbb{Q},\mathbb{P}})^q\big(|\lambda_t|^2+|\alpha^\mathbb{Q}_t|^2\big)dt+L_t,
	\end{align*}
	where $L$ is the local martingale in the Doob-Meyer decomposition of the $\mathbb{P}$-submartingale process $(D^{\mathbb{Q},\mathbb{P}})^q\ln_qD^{\mathbb{Q},\mathbb{P}}$. By Lemma \ref{lemma1}, $(D^{\mathbb{Q},\mathbb{P}})^q\ln_qD^{\mathbb{Q},\mathbb{P}}$ is of class $D$,  and this further implies $L$ is in fact a uniformly integrable martingale by Theorem 1.4.1 of  \cite{KS98}.
	
	The result \eqref{hqt2}  is an immediate consequence of integrating  from  $t$  to $T$, taking the conditional expectations on both sides, and the equality \eqref{hqt1}.   Setting $t=0$,  then \eqref{eq:2}  holds obviously.
	
	To prove equality \eqref{eq:4}, it is enough to show  $H_q(\mathbb{Q}^{min}|\mathbb{P})<+\infty$. In fact, $$H_q(\mathbb{Q}^{min}|\mathbb{P})=\frac{1}{1-q}\Big(1-\mathbb{E}_\mathbb{P}[(\mathcal{E}(-\lambda\cdot W)_T)^q]\Big)$$
	and \begin{align*}
	\mathbb{E}_\mathbb{P}[(\mathcal{E}(-\lambda\cdot W)_T)^q]&=\mathbb{E}_\mathbb{P}\left[(\mathcal{E}(-q\lambda\cdot W)_T)\cdot \exp\left(\frac{q^2-q}{2}\int_0^T|\lambda_s|^2ds\right)\right]\\
	&\leq C_0 \mathbb{E}_\mathbb{P}[(\mathcal{E}(-q\lambda\cdot W)_T)]=C_0,
	\end{align*}
	where the inequality is from the boundedness of $\lambda$ and $C_0$ is a suitable positive constant. We then can get the finiteness of Tsallis relative entropy between $\mathbb{Q}^{min}$ and $\mathbb{P}$.
\hfill$\Box$

\subsection{Proof of Theorem \ref{main}}\label{seca2}

This subsection gives the proof of Theorem \ref{main}.

 \begin{lemma}\label{dual}
	Suppose $\gamma>0$, $q>0$ and $q\neq 1$. For each  $\xi\in \mathcal{L}^{\gamma}_{q}(\mathcal{F}_T, b)$, we have
\begin{align}\label{mainresult1}
Y_t(\xi)=\tilde{F}_t(\xi),   ~~~\forall t\in[0,T],
\end{align}where $Y_{\cdot}(\xi)$ is the solution of BSDE \eqref{bsde}.  Moreover, the essential infimum in \eqref{dualrep} can be achieved.
\end{lemma}
\textbf{Proof.}   By the definition of $\tilde{F}$, for each  $\xi\in \mathcal{L}^{\gamma}_{q}(\mathcal{F}_T, b)$, we only need to show that
	\begin{equation*}\label{dual-y}
	Y_{t}= \mathop{\essinf}_{\theta\in \Theta} Y^{\theta}_{t},  ~~t\in[0,T], ~~\mathbb{P}-a.s.,
	\end{equation*}where $Y$ is the uniqueness solution of BSDE \eqref{bsde}.
	
	For any $\theta\in \Theta$, due to the fact that  $\xi$ is bounded and backward equation \eqref{ty} has a unique solution $Y^{\theta}$ with $\mathbb{E}_{\mathbb{Q}^{\theta}}\big[~\xi+\int_{0}^{T}\frac{|\mu(Y^{\theta}_{s})|}{2\gamma}|\theta_{s}|^{2}ds\big]<\infty$, by martingale representation theorem, then there exists a $\mathbf{R}^{m+n}$-valued predictable process $\bar{\varphi}=(\varphi, \varphi^{\perp})$ with $\int_{0}^{T}|\bar\varphi_{s}|^{2}ds<+\infty$, $\mathbb{Q}^{\theta}$-a.s., such that
	\begin{align}\label{ytheta}
	dY^{\theta}_{s}=-\frac{1}{2\gamma}\mu(Y^{\theta}_{s})\cdot |\theta_{s}|^{2}ds+\varphi_{s}\cdot dW^{-\lambda}_{s}+\varphi_s^{\perp}\cdot dW^{\perp,\theta}_{s},     ~~~~Y^{\theta}_{T}=\xi,
	\end{align}where the process
	\begin{align*}
	\overline{W}_t^\theta=\left(\begin{array}{lll}
	W^{-\lambda}_{t}\\
	W^{\perp,\theta}_{t}
	\end{array}
	\right)=\left(
	\begin{array}{lll}
	W_t+\int_{0}^{t}\lambda_sds\\
	W_t^\perp-\int_{0}^{t}\theta_sds
	\end{array}
	\right),  ~~ 0\leq t\leq T,
	\end{align*} is a $\mathbb{Q}^{\theta}$ Brownian motion on $(\Omega, \mathcal{F}_T)$ by Girsanov theorem.

	On the other hand, BSDE \eqref{bsde} implies that
	\begin{align}\label{yyy1}
	d Y_{s}&=\left(\lambda_s\cdot  Z_s+\frac{\gamma }{2}\cdot\frac{ |Z_{s}^{\perp}|^{2}}{\mu( Y_{s})}\right)ds+Z_{s}\cdot dW_{s}+Z_s^{\perp}\cdot dW_s^{\perp}\nonumber\\
	&=\left(\frac{\gamma }{2}\cdot\frac{ |Z_{s}^{\perp}|^{2}}{\mu( Y_{s})}+\theta_s\cdot Z_s^{\perp}\right)ds+Z_{s}\cdot dW_{s}^{-\lambda}+Z_s^{\perp}\cdot dW_s^{\perp, \theta},~~~~~Y_{T}=\xi.
	\end{align}
	Let's subtract these two equations \eqref{ytheta} and \eqref{yyy1}, then it derives
	\begin{align*}
	d (Y_{s}-Y^{\theta}_{s})=&\left(\frac{\gamma }{2}\cdot\frac{ |Z_{s}^{\perp}|^{2}}{\mu( Y_{s})}+\theta_s\cdot Z_s^{\perp}+\frac{1}{2\gamma}\mu(Y^{\theta}_{s})\cdot |\theta_{s}|^{2}\right)ds\\
	&~~+(Z_{s}-\varphi_{s})\cdot dW_{s}^{-\lambda}+(Z_{s}^{\perp}-\varphi_{s}^{\perp})\cdot dW_{s}^{\perp, \theta}\\
	=&-\frac{\mu(Y_{s})-\mu(Y^{\theta}_{s}) }{2\gamma}|\theta_{s}|^{2}ds+\frac{\mu(Y_{s})}{2\gamma
	}\big|\theta_{s}+\frac{\gamma Z_{s}^{\perp}}{\mu(Y_{s})}\big|^{2}ds\\
	&~~+(Z_{s}-\varphi_{s})\cdot dW_{s}^{-\lambda}+(Z_{s}^{\perp}-\varphi_{s}^{\perp})\cdot dW_{s}^{\perp, \theta}.
	\end{align*}
	From Proposition \ref{bsde:eq2}, we know that $\int_{0}^{\cdot}\overline{Z}_{s}\cdot  d\overline{W}_{s}$ is a BMO martingale under $\mathbb{P}$. Since $\theta\in\Theta$ and boundedness of $\lambda$, by Theorem 3.3 in   \cite{K94}, we also can get that $\int_{0}^{\cdot}\overline{Z}_{s}\cdot d\overline{W}_{s}^{\theta}$ is a BMO martingale under $\mathbb{Q}^{\theta}$.
	
	For $n\geq 1$, define a stopping time sequence, $$\tau_{n}:=\inf\{ r\in[0,T]~|~\int_{0}^{r}|\bar{\varphi}_{s}|^{2}ds\geq n\}\wedge T.$$  For any $t\in[0, \tau_{n}]$, it implies that
	\begin{align}\label{xiangdengtheta}
	Y_{t}-Y^{\theta}_{t}=&\mathbb{E}_{\mathbb{Q}^{\theta}}[Y_{\tau_{n}}-Y^{\theta}_{\tau_{n}}|\mathcal{F}_{t}]+\mathbb{E}_{\mathbb{Q}^{\theta}}\big[\int_{t}^{\tau_{n}} \frac{\mu(Y_{s})-\mu(Y^{\theta}_{s}) }{2\gamma}|\theta_{s}|^{2}ds |\mathcal{F}_{t} \big]\nonumber\\
	&-  \mathbb{E}_{\mathbb{Q}^{\theta}}\big[\int_{t}^{\tau_{n}}  \frac{\mu(Y_{s})}{2\gamma}\big|\theta_{s}+\frac{\gamma Z_{s}^{\perp}}{\mu(Y_{s})}\big|^{2}ds |\mathcal{F}_{t} \big].
	\end{align}Due to the fact that $Y$ is bounded and  $\mu(Y_{s})>0$, $s\in[0,T]$, then
	\begin{align*}
	Y_{t}-Y^{\theta}_{t}\leq \mathbb{E}_{\mathbb{Q}^{\theta}}[Y_{\tau_{n}}-Y^{\theta}_{\tau_{n}}|\mathcal{F}_{t}]+\mathbb{E}_{\mathbb{Q}^{\theta}}\big[\int_{t}^{\tau_{n}} \frac{\mu(Y_{s})-\mu(Y^{\theta}_{s}) }{2\gamma}|\theta_{s}|^{2}ds |\mathcal{F}_{t} \big].
	\end{align*}
	Since the boundedness of $Y$ and $Y^\theta$ and $\theta\in\Theta$, letting $n$ approach infinity, it gives
	\begin{align*}
	Y_{t}-Y^{\theta}_{t}&\leq \mathbb{E}_{\mathbb{Q}^{\theta}}\big[\int_{t}^{T} \frac{\mu(Y_{s})-\mu(Y^{\theta}_{s}) }{2\gamma}|\theta_{s}|^{2}ds |\mathcal{F}_{t} \big]\\
	&=  \mathbb{E}_{\mathbb{Q}^{\theta}}\big[\int_{t}^{T} \frac{q-1}{2q }(Y_{s}-Y^{\theta}_{s})|\theta_{s}|^{2}ds |\mathcal{F}_{t} \big] .
	\end{align*}By Lemma C3 in \cite{SS99}, it implies that
	$$Y_{t}\leq Y^{\theta}_{t}, ~~~~0\leq t\leq T,~~~~~\mathbb{P}-a.s.$$
	Thus, we have $Y_{t}\leq \mathop{\essinf}_{\theta\in \Theta} Y^{\theta}_{t},  ~~t\in[0,T], ~~\mathbb{P}-a.s$.
	
	Finally, from the procedure of above proof, if we
	choose $$\theta^{*}:=-\frac{\gamma Z^\perp}{\mu(Y)}$$ in \eqref{xiangdengtheta} and verify $\theta^*\in\Theta$, then it concludes $Y=Y^{\theta^*}$.

	In fact, since $\mu(Y)$ is bounded, then   $\int_{0}^{\cdot}\theta^{*}_{s}\cdot dW_{s}^{\perp}$ is obviously a BMO($\mathbb{P}$) martingale by Proposition \ref{bsde:eq2}.
	Backward equation \eqref{ty} has at most one solution by Lemma C3 in \cite{SS99}. For the given $\theta^*$, we can show that $Y$ is the solution of \eqref{ty}.  Besides, \begin{align*}
	\mathbb{E}_{\mathbb{Q}^{\theta^{*}}}\big[\int_{0}^{T}|\mu(Y_{s})|\cdot |\theta^*_s|^2ds\big]&= \mathbb{E}_{\mathbb{Q}^{\theta^{*}}}\big[\int_{0}^{T}\frac{\gamma^2 |Z_{s}^\perp|^{2}}{2\mu(Y_{s})}ds\big]\\
	&\leq C_1 \mathbb{E}_{\mathbb{Q}^{\theta^{*}}}\big[\int_{0}^{T}|Z_{s}^\perp|^{2}ds\big]<+\infty,
	\end{align*}where $C_1$ is a suitable positive constant.

	Hence, it yields the assertion, i.e., \eqref{mainresult1} holds.  \hfill$\Box$

\begin{remark}
	Lemma \ref{dual} shows that the solution of BSDE \eqref{bsde} equals to the essential infimum of backward equations \eqref{ty}  under a family of equivalent martingale measures $\mathcal{M}^{\Theta}$. 
It is a robust dual representation result for BSDE \eqref{bsde}. Essentially,  the key tool to obtain this robust dual representation depends on the Legendre transform of the concave generator of BSDE \eqref{bsde}.
	
	For the robust representation results of BSDEs with concave or convex generators, the reader can refer to \cite{EPQ97} with Lipschitz generators,  \cite{BK09},  \cite{LQ03} or  \cite{ER09} with quadratic growth generators.
\end{remark}

\begin{lemma}\label{dual-tsallis}
	Suppose $\gamma>0$, $q>0$ and $q\neq 1$. For any $\xi \in\mathcal{L}^{\gamma}_{q}(\mathcal{F}_T, b)$, for each $t\in[0,T]$,  we have
	\begin{equation}\label{dualrep11}
	Y_t(\xi)=F_t(\xi)= \mathop{\essinf}_{\mathbb{Q}\in \mathcal{M}_f} \mathbb{E}_{\mathbb{Q}^{min}}\left[(D_{t,T}^{\mathbb{Q},\mathbb{Q}^{min}})^q~\xi~+\frac{q}{2 \gamma}\int_{t}^{T}(D_{t,s}^{\mathbb{Q},\mathbb{Q}^{min}})^q|\alpha^{\mathbb{Q}}_s|^2ds~\big|~\mathcal{F}_t\right].
	\end{equation}Moreover, the essential infimum  can be attained.
	In particular,
	\begin{align}
	Y_0(\xi)=F(\xi)&= \mathop{\inf}_{\mathbb{Q}\in \mathcal{M}_f} \mathbb{E}_{\mathbb{Q}^{min}}\left[(D_{T}^{\mathbb{Q},\mathbb{Q}^{min}})^q~\xi~+\frac{q}{2 \gamma}\int_{0}^{T}(D_{s}^{\mathbb{Q},\mathbb{Q}^{min}})^q|\alpha^{\mathbb{Q}}_s|^2ds\right] \label{dualrep21}\\
	&= \mathop{\inf}_{\mathbb{Q}\in \mathcal{M}_f} \left( \mathbb{E}_{\mathbb{Q}^{min}}\big[(D_{T}^{\mathbb{Q},\mathbb{Q}^{min}})^q~\xi~\big]+\frac{1}{ \gamma} H_q(\mathbb{Q}|\mathbb{Q}^{min}) \label{dualrep22}\right).
	\end{align}

\end{lemma}
\textbf{Proof.}  For any $\mathbb{Q}\in \mathcal{M}_f$,
	since
	$$dD_s^{\mathbb{Q},\mathbb{Q}^{min}}=D_s^{\mathbb{Q},\mathbb{Q}^{min}} \alpha_s^\mathbb{Q}\cdot dW_s^\perp,~~ D_0^{\mathbb{Q},\mathbb{Q}^{min}}=1,$$
	applying It\^{o}'s formula to $(D^{\mathbb{Q},\mathbb{Q}^{min}})^q$, it derives that
	\begin{align*}
	d(D_s^{\mathbb{Q},\mathbb{Q}^{min}})^q&=q(D_s^{\mathbb{Q},\mathbb{Q}^{min}})^q\alpha_s^\mathbb{Q}\cdot dW_s^\perp+\frac{q(q-1)}{2} (D_s^{\mathbb{Q},\mathbb{Q}^{min}})^q|\alpha_s^\mathbb{Q}|^2ds.
	\end{align*}
	From the finiteness of Tsallis relative entropy between $\mathbb{Q}$ and $\mathbb{Q}^{min}$, we get
	$(D_T^{\mathbb{Q},\mathbb{Q}^{min}})^q\in L^1(\mathbb{Q}^{min})$. For $q>1$, by Jensen's inequality we know that the process $(D^{\mathbb{Q},\mathbb{Q}^{min}})^q$ is a uniformly integrable submartingale under  $\mathbb{Q}^{min}$. Similarly, for $0<q<1$, $(D^{\mathbb{Q},\mathbb{Q}^{min}})^q$ is a uniformly integrable supermartingale under  $\mathbb{Q}^{min}$.
	Thus, for $q>0$ and $q\neq 1$ we know that $(D^{\mathbb{Q},\mathbb{Q}^{min}})^q$ is of class $D$.

	On the other hand, BSDE \eqref{bsde} can be rewritten as follows,
	
	\begin{equation}\label{bianlab}
	d Y_{s}=\frac{\gamma }{2}\cdot\frac{ |Z_{s}^{\perp}|^{2}}{\mu( Y_{s})}ds+Z_{s}\cdot dW_{s}^{-\lambda}+Z_s^{\perp}\cdot dW_s^{\perp},~~Y_T=\xi,
	\end{equation}
	where  $dW_s^{-\lambda}=dW_s+\lambda_sds$. By Girsanov theorem, we know that   $\overline{W}^{min}=(W^{-\lambda}, W^{\perp})$ is a Brownian motion under minimal martingale measure $\mathbb{Q}^{min}$, and $\langle W^{-\lambda}, W^{\perp}\rangle=0$. Since $Y$ is boundedness and $\int_0^\cdot \overline{Z}_s\cdot d\overline{W}_s$ is a BMO martingale under $\mathbb{P}$, then, by Theorem 3.3 in \cite{K94}, $\int_0^\cdot \overline{Z}_s\cdot d\overline{W}_s^{min}$ is a BMO martingale under $\mathbb{Q}^{min}$.

	Using It\^{o}'s formula to $Y(D^{\mathbb{Q},\mathbb{Q}^{min}})^q$, we get that
	\begin{align*}
	dY_s(D_s^{\mathbb{Q},\mathbb{Q}^{min}})^q&=Y_sd(D_s^{\mathbb{Q},\mathbb{Q}^{min}})^q+(D_s^{\mathbb{Q},\mathbb{Q}^{min}})^qdY_s+dY_sd(D_s^{\mathbb{Q},\mathbb{Q}^{min}})^q\\
	&=(D_s^{\mathbb{Q},\mathbb{Q}^{min}})^q\left(\frac{q(q-1)}{2} Y_s \cdot|\alpha_s^\mathbb{Q}|^2+\frac{\gamma }{2}\cdot\frac{ |Z_{s}^{\perp}|^{2}}{\mu( Y_{s})}+q\alpha_s^\mathbb{Q}\cdot Z^{\perp}_s\right)ds\\
	&~~~~+(D_s^{\mathbb{Q},\mathbb{Q}^{min}})^q\left(Z_{s}\cdot dW_{s}^{-\lambda}+(Z_s^{\perp}+qY_s\alpha_s^\mathbb{Q})\cdot dW_s^\perp\right).
	\end{align*}
	The stochastic integral term $L_\cdot=\int_0^\cdot(D_s^{\mathbb{Q},\mathbb{Q}^{min}})^q\left(Z_{s}\cdot dW_{s}^{-\lambda}+(Z_s^{\perp}+qY_s\alpha_s^\mathbb{Q})\cdot dW_s^\perp\right)$ is a local martingale under $\mathbb{Q}^{min}$ by It\^{o}'s formula. Let $\tau_n$ be a sequence of stopping times converging to $T$ such that $\tau_n\geq t$ and $L^{\tau_n}$ is a true martingale. Then we get
	
	\begin{equation}\label{yt-tiaojian}
	Y_t=\mathbb{E}_{\mathbb{Q}^{min}}\left[Y_{\tau_n}(D_{t,\tau_n}^{\mathbb{Q},\mathbb{Q}^{min}})^q~|~\mathcal{F}_t~\right]-\mathbb{E}_{\mathbb{Q}^{min}}\left[\int_t^{\tau_n}(D_{t,s}^{\mathbb{Q},\mathbb{Q}^{min}})^q\Gamma_sds~\big|~\mathcal{F}_t~\right],
	\end{equation}
	where $$
	\Gamma_s:=\frac{q(q-1)}{2} Y_s \cdot|\alpha_s^\mathbb{Q}|^2+\frac{\gamma }{2}\cdot\frac{ |Z_{s}^{\perp}|^{2}}{\mu( Y_{s})}+q\alpha_s^\mathbb{Q}\cdot Z^{\perp}_s.
	$$By the definition of $\mu(\cdot)$, it implies that
	\begin{align*}
	\Gamma_s&=\frac{q}{2\gamma} (q-1)\gamma Y_s \cdot|\alpha_s^\mathbb{Q}|^2+\frac{\gamma }{2}\cdot\frac{ |Z_{s}^{\perp}|^{2}}{\mu( Y_{s})}+q\alpha_s^\mathbb{Q}\cdot Z^{\perp}_s\\
	&=\frac{q}{2\gamma}(\mu(Y_s)q-1)\cdot|\alpha_s^\mathbb{Q}|^2+\frac{\gamma }{2}\cdot\frac{ |Z_{s}^{\perp}|^{2}}{\mu( Y_{s})}+q\alpha_s^\mathbb{Q}\cdot Z^{\perp}_s\\
	&=\frac{\mu(Y_s)}{2\gamma}|q\alpha_s^\mathbb{Q}+\frac{\gamma Z_s^{\perp}}{\mu(Y_s)}|^2-\frac{q}{2\gamma}\cdot|\alpha_s^\mathbb{Q}|^2.
	\end{align*}
	Putting the $\Gamma$ into \eqref{yt-tiaojian}, we get

	\begin{align}\label{yt-tiaojian2}
	Y_t&=\mathbb{E}_{\mathbb{Q}^{min}}\left[Y_{\tau_n}(D_{t,\tau_n}^{\mathbb{Q},\mathbb{Q}^{min}})^q~|~\mathcal{F}_t~\right]+\frac{q}{2\gamma}\mathbb{E}_{\mathbb{Q}^{min}}\left[\int_t^{\tau_n}(D_{t,s}^{\mathbb{Q},\mathbb{Q}^{min}})^q\cdot|\alpha_s^\mathbb{Q}|^2ds~\big|~\mathcal{F}_t~\right]\nonumber\\
	&~~~~-\frac{1}{2\gamma}\mathbb{E}_{\mathbb{Q}^{min}}\left[\int_t^{\tau_n}(D_{t,s}^{\mathbb{Q},\mathbb{Q}^{min}})^q\mu(Y_s)|q\alpha_s^\mathbb{Q}+\frac{\gamma Z_s^{\perp}}{\mu(Y_s)}|^2ds~\big|~\mathcal{F}_t~\right].
	\end{align}
	Since $(D^{\mathbb{Q},\mathbb{Q}^{min}})^q$ is of class $D$, $Y_\cdot$ and $\mu(Y_\cdot)$ are bounded, sending $n$ to infinity, by  monotone convergence theorem, which yields that
	
	\begin{align}\label{yt-tiaojian3}
	Y_t&=\mathbb{E}_{\mathbb{Q}^{min}}\left[(D_{t,T}^{\mathbb{Q},\mathbb{Q}^{min}})^q\xi~|~\mathcal{F}_t~\right]+\frac{q}{2\gamma}\mathbb{E}_{\mathbb{Q}^{min}}\left[\int_t^{T}(D_{t,s}^{\mathbb{Q},\mathbb{Q}^{min}})^q\cdot|\alpha_s^\mathbb{Q}|^2ds~\big|~\mathcal{F}_t~\right]\nonumber\\
	&~~~~-\frac{1}{2\gamma}\mathbb{E}_{\mathbb{Q}^{min}}\left[\int_t^{T}(D_{t,s}^{\mathbb{Q},\mathbb{Q}^{min}})^q\mu(Y_s)\cdot|q\alpha_s^\mathbb{Q}+\frac{\gamma Z_s^{\perp}}{\mu(Y_s)}|^2ds~\big|~\mathcal{F}_t~\right].\nonumber
	\end{align}Hence,
	$$Y_t\leq \mathop{\essinf}_{\mathbb{Q}\in \mathcal{M}_f} \mathbb{E}_{\mathbb{Q}^{min}}\left[(D_{t,T}^{\mathbb{Q},\mathbb{Q}^{min}})^q\xi+\frac{q}{2\gamma}\int_t^{T}(D_{t,s}^{\mathbb{Q},\mathbb{Q}^{min}})^q\cdot|\alpha_s^\mathbb{Q}|^2ds~\big|~\mathcal{F}_t~\right].
	$$
	
	Moreover, the essential infimum  is actually achieved if we can show
	$\mathbb{Q}^*\in \mathcal{M}_f$, where
	\begin{equation}\label{alpha*}
	\alpha^{\mathbb{Q}^*}:=-\frac{\gamma Z^{\perp}}{q\mu(Y)}.
	\end{equation}Since
	$\int_0^\cdot \overline{Z}_s\cdot d\overline{W}_s^{min}$ is a BMO martingale under $\mathbb{Q}^{min}$ and $\mu(Y)$ is bounded from below,  then $\int_0^\cdot \alpha_s^{\mathbb{Q}^*}\cdot d{W}_s^{\perp}$  is also a BMO martingale under $\mathbb{Q}^{min}$. By Theorem 2.3 in  \cite{K94}, we get $\mathcal{E}(\alpha^{\mathbb{Q}^*}\cdot {W}^{\perp})$ is a uniformly integrable martingale under $\mathbb{Q}^{min}$. Thus, we can define
	$$\frac{d\mathbb{Q}^*}{d\mathbb{Q}^{min}}:=\mathcal{E}(\alpha^{\mathbb{Q}^*}\cdot {W}^{\perp})_T=D_T^{\mathbb{Q}^*,\mathbb{Q}^{min}}, $$and $\mathbb{Q}^*\in \mathcal{M}$.
	
	By the following Lemma \ref{zhishuyangq}, we know that $H_q(\mathbb{Q}^*~|~\mathbb{Q}^{min})<+\infty$.  To sum up, \eqref{dualrep11} is true. Setting $t=0$, then \eqref{dualrep21} and \eqref{dualrep22} can be easily obtained by \eqref{eq:qmin} in Corollary \ref{qqmin}.   	   		\hfill$\Box$

\begin{lemma}\label{zhishuyangq}
	Suppose $\gamma>0$, $q>0$ and $q\neq 1$. Suppose   $\alpha^{\mathbb{Q}^*}=-\frac{\gamma Z^{\perp}}{q\mu(Y)}$. We have
	
	\begin{equation}\label{holder}
	\mathbb{E}_{\mathbb{Q}^{min}}\left[(D_T^{\mathbb{Q}^*,\mathbb{Q}^{min}})^q\right]=\mathbb{E}_{\mathbb{Q}^{min}}\left[\mathcal{E}(\alpha^{\mathbb{Q}^*}\cdot {W}^{\perp})_T^q\right]<+\infty,
	\end{equation}
	which means  $H_q(\mathbb{Q}^*~|~\mathbb{Q}^{min})<+\infty$.
\end{lemma}
\textbf{Proof.} 
	When $0<q<1$, it holds obviously by H\"{o}lder inequality.
	
	Next, we are going to prove the situation of  $q>1$. This case is not trivial.   Using It\^{o}'s formula to $\ln \mu(Y)$, it implies that
	\begin{align*}
	d\ln \mu(Y_s)&=\frac{(q-1)\gamma }{q\mu(Y_s)}dY_s-\frac{1}{2}\frac{(q-1)^2\gamma^2}{q^2}\frac{1}{\mu^2(Y_s)}\langle dY_s\rangle^2\\
	&=\frac{(q-1)\gamma }{q}\frac{Z_s}{\mu(Y_s)}\cdot dW_s^{-\lambda}-\frac{1}{2}\frac{(q-1)^2\gamma^2}{q^2}\frac{|Z_s|^2}{\mu^2(Y_s)}ds\\
	&~~~~-(q-1)\alpha_s^{\mathbb{Q}^*}\cdot dW_s^\perp+\frac{q-1}{2}|\alpha_s^{\mathbb{Q}^*}|^2ds,
	\end{align*}
	where $W^{-\lambda}$ is defined the same as in \eqref{bianlab}.
	
	Integrating both sides from zero to $T$, it derives that
	\begin{align*}
	&(q-1)\left(\int_0^T\alpha_s^{\mathbb{Q}^*}\cdot dW_s^\perp-\frac{1}{2}\int_0^T|\alpha_s^{\mathbb{Q}^*}|^2ds\right)\\
	=&\ln\frac{\mu(Y_0)}{\mu(\xi)}+\int_0^T\frac{(q-1)\gamma }{q}\frac{Z_s}{\mu(Y_s)}\cdot dW_s^{-\lambda}-\int_0^T\frac{1}{2}\frac{(q-1)^2\gamma^2}{q^2}\frac{|Z_s|^2}{\mu^2(Y_s)}ds.
	\end{align*}
	Taking the exponent on both sides of the above equality, it implies
	
	$$
	\mathcal{E}(\alpha^{\mathbb{Q}^*}\cdot {W}^{\perp})_T^{q-1}=\frac{\mu(Y_0)}{\mu(\xi)}\mathcal{E}\left(\frac{(q-1)\gamma }{q}\frac{Z}{\mu(Y)}\cdot W^{-\lambda}\right)_T.
	$$Therefore, we get that
	$$\mathcal{E}(\alpha^{\mathbb{Q}^*}\cdot {W}^{\perp})_T^{q}=\frac{\mu(Y_0)}{\mu(\xi)}\mathcal{E}\left(\frac{(q-1)\gamma }{q}\frac{Z}{\mu(Y)}\cdot W^{-\lambda}\right)_T\mathcal{E}(\alpha^{\mathbb{Q}^*}\cdot {W}^{\perp})_T.$$

	Since $\int_0^\cdot\frac{(q-1)\gamma }{q}\frac{Z_s}{\mu(Y_s)}\cdot dW_s^{-\lambda}$ and
	$\int_0^{\cdot} \alpha_s^{\mathbb{Q}^*}d{W}_s^{\perp}$
	are both the BMO martingales under $\mathbb{Q}^{min}$.  By Theorem 2.3 in  \cite{K94},
	we get $\mathcal{E}\left(\frac{(q-1)\gamma }{q}\frac{Z}{\mu(Y)}\cdot W^{-\lambda}+\alpha^{\mathbb{Q}^*}\cdot {W}^{\perp}\right)$ is a uniformly integrable martingale under $\mathbb{Q}^{min}$. Due to the fact that $\mu(Y)$ is bounded, we can get that
	\begin{align*}
	\mathbb{E}_{\mathbb{Q}^{min}}\left[\mathcal{E}(\alpha^{\mathbb{Q}^*}\cdot {W}^{\perp})_T^q\right]&\leq C_2\mathbb{E}_{\mathbb{Q}^{min}}\left[\mathcal{E}\left(\frac{(q-1)\gamma }{q}\frac{Z}{\mu(Y)}\cdot W^{-\lambda}+\alpha^{\mathbb{Q}^*}\cdot {W}^{\perp}\right)_T\right]\\&=C_2 <+\infty,\end{align*}where $C_2$ is a suitable positive constant. 	
\hfill$\Box$

\begin{remark}
	
	In the situation of $q>1$,  by Theorem 3.4 in  \cite{K94} and BMO martingale property of $\int_0^\cdot \alpha_s^{\mathbb{Q}^*}d{W}_s^{\perp}$ under $\mathbb{Q}^{min}$, then we get $\mathcal{E}(\alpha^{\mathbb{Q}^*}\cdot {W}^{\perp})$ satisfies the reverse  H\"{o}lder inequality ($R_p$) for some $p>1$, i.e.,
	
	$$\mathbb{E}_{\mathbb{Q}^{min}}\left[\mathcal{E}(\alpha^{\mathbb{Q}^*}\cdot {W}^{\perp})_T^p\right]\leq \tilde{C}_p\mathcal{E}(\alpha^{\mathbb{Q}^*}\cdot {W}^{\perp})_0^p=\tilde{C}_p<+\infty,$$where $\tilde{C}_p$ is a positive constant. But,  we can't simply get the magnitude relationship between  $q$ and $p$.   If $p\geq q$, then \eqref{holder} is true by H\"{o}lder inequality again. However, if
	$p<q$, we can not get the result directly.  Lemma \ref{zhishuyangq}  essentially depends on the structure of BSDE \eqref{bsde}.
\end{remark}

\noindent{\bf{Proof of Theorem \ref{main}.}}  Theorem \ref{main} is immediately obtained by combining Lemma \ref{dual} and Lemma \ref{dual-tsallis}. \hfill$\Box$

\subsection{Proof of Proposition \ref{dual-tsallis-z}}

\noindent\textbf{Proof.}  For any $\mathbb{Q}\in \mathcal{Q}_f$,  noticing that
	$$dD^{\mathbb{Q},\mathbb{Q}^{min}}_s=D^{\mathbb{Q},\mathbb{Q}^{min}}_s(\beta_s^{\mathbb{Q}}\cdot dW^{-\lambda}_s+\alpha^\mathbb{Q}_s\cdot dW^\perp_s),~~D^{\mathbb{Q},\mathbb{Q}^{min}}_0=1$$ and $$d y_{s}=\frac{\gamma }{2}\cdot\frac{ |z_{s}|^{2}+|z_{s}^{\perp}|^{2}}{\mu( y_{s})}ds+z_{s}\cdot dW_{s}^{-\lambda}+z_s^{\perp}\cdot dW_s^{\perp}, ~~y_T=\xi,$$ where $\beta^{\mathbb{Q}}$ and $\alpha^{\mathbb{Q}}$ are determined by the martingale representation theorem of Brownian motion.

	By the similar arguments in  Lemma \ref{dual-tsallis}, we can get
	
	\begin{align}\label{yt-tiaojian34}
	y_t&=\mathbb{E}_{\mathbb{Q}^{min}}\left[(D_{t,T}^{\mathbb{Q},\mathbb{Q}^{min}})^q\xi~|~\mathcal{F}_t~\right]+\frac{q}{2\gamma}\mathbb{E}_{\mathbb{Q}^{min}}\left[\int_t^{T}(D_{t,s}^{\mathbb{Q},\mathbb{Q}^{min}})^q\cdot(|\beta_s^{\mathbb{Q}}|^2+|\alpha^{\mathbb{Q}}_s|^2) ds~\big|~\mathcal{F}_t~\right]\nonumber\\
	&~~~~-\frac{1}{2\gamma}\mathbb{E}_{\mathbb{Q}^{min}}\left[\int_t^{T}(D_{t,s}^{\mathbb{Q},\mathbb{Q}^{min}})^q\mu(y_s)\cdot\left(|q\beta_s^\mathbb{Q}+\frac{\gamma z_s}{\mu(y_s)}|^2+|q\alpha_s^\mathbb{Q}+\frac{\gamma z_s^{\perp}}{\mu(y_s)}|^2\right)ds~\big|~\mathcal{F}_t~\right].\nonumber
	\end{align}Therefore,
	$$y_t\leq \mathop{\essinf}_{\mathbb{Q}\in \mathcal{Q}_f} \mathbb{E}_{\mathbb{Q}^{min}}\left[(D_{t,T}^{\mathbb{Q},\mathbb{Q}^{min}})^q\xi+\frac{q}{2\gamma}\int_t^{T}(D_{t,s}^{\mathbb{Q},\mathbb{Q}^{min}})^q\cdot(|\beta_s^{\mathbb{Q}}|^2+|\alpha^{\mathbb{Q}}_s|^2)ds~\big|~\mathcal{F}_t~\right].
	$$
	Besides, the essential infimum  is actually achieved if we can show
	$\mathbb{Q}^*\in \mathcal{Q}_f$, where
	\begin{equation}\label{alpha*1}
	\alpha^{\mathbb{Q}^*}:=-\frac{\gamma z^{\perp}}{q\mu(y)}~~~\textrm{and}~~~\beta^{\mathbb{Q}^*}:=-\frac{\gamma z}{q\mu(y)}.
	\end{equation}
	
	In fact, since
	$\int_0^\cdot \overline{z}_s\cdot d\overline{W}_s^{min}$ is a BMO martingale under $\mathbb{Q}^{min}$ and $\mu(y)$ is bounded from below,   by Theorem 2.3 in  \cite{K94}, we can get $\mathcal{E}(\beta^{\mathbb{Q}^*}\cdot W^{-\lambda}+\alpha^{\mathbb{Q}^*}\cdot W^\perp)$ is a uniformly integrable martingale under $\mathbb{Q}^{min}$, which implies  $\mathbb{Q}^*\in \mathcal{Q}$.
	
	Next, we claim $\mathbb{Q}^*\in \mathcal{Q}_f$, i.e., $H_q(\mathbb{Q}^*|\mathbb{Q}^{min})<+\infty$. When $q<1$, it is obvious. For $q>1$, applying
	It\^{o}'s formula to $\frac{q}{1-q}\ln \mu(y)$, it derives that
	\begin{align*}
	\frac{q}{1-q}d\ln \mu(y_s)
	&=q\left(\beta_s^{\mathbb{Q}^*}\cdot dW^{-\lambda}_s+\alpha_s^{\mathbb{Q}^*}\cdot dW_s^\perp-\frac{1}{2}(|\beta_s^{\mathbb{Q}^*}|^2+|\alpha_s^{\mathbb{Q}^*}|^2)ds\right),
	\end{align*}
	which leads to
	\begin{align*}\mathbb{E}_{\mathbb{Q}^{min}}\big[(D_T^{\mathbb{Q}^*,\mathbb{Q}^{min}})^q\big]&=\mathbb{E}_{\mathbb{Q}^{min}}\big[\mathcal{E}(\beta^{\mathbb{Q}^*}\cdot W^{-\lambda}+\alpha^{\mathbb{Q}^*}\cdot W^\perp)_T^q\big]\\
	&=\mathbb{E}_{\mathbb{Q}^{min}}\big[\left(\frac{\mu(\xi)}{\mu(y_0)}\right)^{q/(1-q)}\big]<+\infty.\end{align*}
	For any $t\in[0,T]$, applying Theorem \ref{shang1} to $H_{q,t}(\mathbb{Q}|\mathbb{Q}^{min})$, it completes the proof.  \hfill$\Box$

\subsection{Proof of Proposition \ref{jixian} }

\noindent\textbf{Proof.}  From Proposition \ref{property3} (i), we know that for each  $t\in[0,T]$ and  each bounded $\xi\in\mathop{\bigcap}_{\gamma>0}\mathcal{L}_{q}^{\gamma}(\mathcal{F}_T, b)$, $F_t(\xi,\gamma)$ is decreasing with respect to $\gamma$.
	
By the definition of $F$,  for any $\gamma>0$,
	we obviously get
	$$F_t(\xi,\gamma)\geq \mathop{\essinf}_{\mathbb{Q}\in \mathcal{M}_f} \mathbb{E}_{\mathbb{Q}^{min}}\left[(D_{t,T}^{\mathbb{Q},\mathbb{Q}^{min}})^q~\xi~\big|~\mathcal{F}_t\right].$$
	Thus, it implies 	$$\lim_{\gamma\rightarrow +\infty}F_t(\xi,\gamma)\geq\mathop{\essinf}_{\mathbb{Q}\in \mathcal{M}_f} \mathbb{E}_{\mathbb{Q}^{min}}\left[(D_{t,T}^{\mathbb{Q},\mathbb{Q}^{min}})^q~\xi~\big|~\mathcal{F}_t\right].$$
	
	On the other hand, for any $\mathbb{Q}\in \mathcal{M}_f$, for any $\gamma>0$, we have
	\begin{align*}
	&F_t(\xi,\gamma)\\
\leq ~&\mathbb{E}_{\mathbb{Q}^{min}}\left[(D_{t,T}^{\mathbb{Q},\mathbb{Q}^{min}})^q \xi~ \big|~\mathcal{F}_t\right]+\frac{1}{\gamma}H_{q,t}(\mathbb{Q}| \mathbb{Q}^{min})\\
	=~&\frac{1}{\gamma} \frac{1}{(D_{t}^{\mathbb{Q},\mathbb{Q}^{min}})^q}\left(\mathbb{E}_{\mathbb{Q}^{min}}\Big[(D_{T}^{\mathbb{Q},\mathbb{Q}^{min}})^q\ln_qD_{T}^{\mathbb{Q},\mathbb{Q}^{min}}~\big|~\mathcal{F}_{t}\Big]-(D_{t}^{\mathbb{Q},\mathbb{Q}^{min}})^q\ln_qD_{t}^{\mathbb{Q},\mathbb{Q}^{min}}\right)\\
&+\mathbb{E}_{\mathbb{Q}^{min}}\left[(D_{t,T}^{\mathbb{Q},\mathbb{Q}^{min}})^q \xi ~\big|~\mathcal{F}_{t}\right],
	\end{align*}where the last equality follows from  \eqref{hqt1} in Theorem \ref{shang1}.  
	Sending $\gamma$ to infinity, and taking the infimum with
	$\mathbb{Q}$ on $\mathcal{M}_{f}$,  \eqref{gamma1} holds.

By Proposition \ref{jixian23}, for each $\gamma>0$, we have 
$$
	CE_{q}(\xi|\mathcal{F}_t)	\leq F_t(\xi, \gamma)\leq \mathbb{E}_{\mathbb{Q}^{min}}[\xi\big|\mathcal{F}_t].
$$
In order to prove \eqref{gamma0}, we only need to show that for each $\xi\in\mathop{\bigcap}_{\gamma>0}\mathcal{L}_{q}^{\gamma}(\mathcal{F}_T, b)$,
\begin{equation}\label{ceqgamma}
\lim_{\gamma\rightarrow 0}-\frac{1}{\gamma}\ln_{q}\mathbb{E}_{\mathbb{Q}^{min}}[\exp_q(-\gamma \xi)|\mathcal{F}_t]=\mathbb{E}_{\mathbb{Q}^{min}}[\xi\big|\mathcal{F}_t],~~~~\forall t\in[0,T].
\end{equation}
In fact, for each $t\in[0,T]$,  setting 
$$h(\gamma)=\mathbb{E}_{\mathbb{Q}^{min}}[\exp_q(-\gamma \xi)|\mathcal{F}_t].$$
Then $h(\gamma)\stackrel{a.s.}{\longrightarrow} 1$ when $\gamma$ goes to zero.  Since $\xi$ is bounded,  we have that  $$h'(\gamma)=-\mathbb{E}_{\mathbb{Q}^{min}}[\exp^q_q(-\gamma \xi)\xi|\mathcal{F}_t],  ~~~ \mathbb{P}-a.s.$$
Therefore,  it implies that 
\begin{align*}
&\lim_{\gamma\rightarrow 0}-\frac{1}{\gamma}\ln_{q}\mathbb{E}_{\mathbb{Q}^{min}}[\exp_q(-\gamma \xi)|\mathcal{F}_t]\\
=&\lim_{\gamma\rightarrow 0}\frac{1-h(\gamma)^{1-q}}{\gamma (1-q)}\\
=&\lim_{\gamma\rightarrow 0}-\frac{h'(\gamma)}{(h(\gamma))^{q}}\\
=&\mathbb{E}_{\mathbb{Q}^{min}}[\xi\big|\mathcal{F}_t].
\end{align*}
\hfill$\Box$

\subsection{Proof of Proposition \ref{bsde:eq2}} 

Suppose $\gamma>0$, $q>0$ and $q\neq 1$. Define
\begin{align*}
\Lambda:=\left\{
\begin{array}{lll}
x<\frac{1}{(1-q)\gamma}~,~~~  0<q<1,\\
x>\frac{1}{(1-q)\gamma}~,~~~ q>1.
\end{array}
\right.
\end{align*}
We denote $f(y):=\frac{\gamma}{2\mu(y)}=\frac{\gamma q}{2}\exp_q^{q-1}(-\gamma y)$, $y\in \Lambda$. Then $f:\Lambda\rightarrow \mathbf{R}$ is locally integrable.

Define $W_{\cdot}^{-\lambda}:=W_\cdot+\int_0^\cdot\lambda_sds$, we know that   $\overline{W}^{min}=(W^{-\lambda}, W^{\perp})$ is a Brownian motion under minimal martingale measure $\mathbb{Q}^{min}$ and $\langle W^{-\lambda}, W^{\perp}\rangle=0$. Obviously, Proposition \ref{bsde:eq2} is  equivalent to the following proposition by Girsanov transformation, see Theorem 3.3 in \cite{K94}.

\begin{proposition}\label{bsde:dengjia} Suppose $\gamma>0$, $q>0$ and $q\neq 1$. For any  $\xi\in \mathcal{L}_{q}^{\gamma}(\mathcal{F}_T, b)$, then the following BSDE
	\begin{align}\label{bsdedengjia}
	Y_{t}&=\xi-\int_t^Tf(Y_s)\cdot |Z_{s}^{\perp}|^{2}ds-\int_t^TZ_{s}\cdot dW^{-\lambda}_{s}-\int_t^TZ_s^{\perp}\cdot dW_s^{\perp},~~~t\in[0,T],
	\end{align}	
	admits a unique solution $(Y, \overline{Z})=(Y, Z, Z^{\perp})$ in which $\int_{0}^{\cdot}\overline{Z}_{s}\cdot d\overline{W}^{min}_s$ is a BMO($\mathbb{Q}^{min}$) martingale, and $Y$ is continuous and bounded, specifically, for each $t\in[0,T]$, $Y_t\in \mathcal{L}_{q}^{\gamma}(\mathcal{F}_t, b)$. 	
\end{proposition}
\textbf{Proof.}
	For any $\xi\in \mathcal{L}_{q}^{\gamma}(\mathcal{F}_T, b)$, then there exist two positive constants $m_1$ and $m_2$ such that  $0<m_{1}\leq\exp_{q}(-\gamma \xi)\leq m_{2}.$ Therefore,
	$$\tilde{m}_2:=-\frac{1}{\gamma}\ln_qm_2\leq \xi\leq -\frac{1}{\gamma}\ln_qm_1=:\tilde{m}_1.$$Since $\ln_qm_1>-\frac{1}{1-q}$ if $0<q<1$, and $\ln_qm_2< -\frac{1}{1-q}$ if $q>1$, it implies that
	$$\xi\in [\tilde{m}_2, \tilde{m}_1]\subseteq \Lambda.$$
	Moreover, for any $y\in \Lambda$, $\bar{z}=(z,z^\perp)\in \mathbf{R}^{m+n}$, $$|F_{\Lambda}(y,z^\perp)|=f(y)|z^{\perp}|^2\leq f(y)|\bar{z}|^2,$$
	then, by Proposition 4.4 in \cite{ZZF21},  there exists a solution $(Y, \overline{Z})=(Y, Z, Z^{\perp})$ in which $\int_{0}^{\cdot}\overline{Z}_{s}\cdot d\overline{W}^{min}_s$ is a BMO($\mathbb{Q}^{min}$) martingale, and $Y$ is continuous and bounded, specifically, for each $t\in[0,T]$, $Y_t\in [\tilde{m}_2, \tilde{m}_1]$ or $Y_t\in \mathcal{L}_{q}^{\gamma}(\mathcal{F}_t, b)$.

	Besides, using Theorem A1 in \cite{BEO17}, in fact, there exist a maximal solution and a minimal solution,  in  $[\tilde{m}_2, \tilde{m}_1]$, for BSDE \eqref{bsdedengjia}. Suppose $(Y^L, \bar{Z}^L)$ and $(Y^U, \bar{Z}^U)$ are minimal solution and maximal solution for BSDE \eqref{bsdedengjia} respectively.

Due to the fact that the generator $F_{\Lambda}(y,z^\perp)=-f(y)|z^{\perp}|^2$ is concave in $(y,z^\perp)$,  motivated by the $\theta$-method in  \cite{BH08} dealing with the convex generator, we can derive the uniqueness of the solution to BSDE \eqref{bsdedengjia} by some subtle transformations. The main difficulty is that we have to be careful about the domain of the function $F_{\Lambda}$.

	Case (i): $0<q<1$.  For any $\theta\in(0,1)$, setting
	\begin{align*}
	&\delta_\theta Y=Y^L-\theta Y^U,
	\delta_\theta Z=Z^L-\theta Z^U, \delta_\theta Z^{\perp}=Z^{\perp,L}-\theta Z^{\perp,U},\\
	&\delta_\theta F_{\Lambda}=F_{\Lambda}(Y^L,Z^{\perp,L})-\theta F_{\Lambda}\big( Y^U, Z^{\perp,U}).
	\end{align*}
	We have that
	$$\frac{\delta_\theta Y}{1-\theta}=\frac{\theta}{1-\theta}(Y^L-Y^U)+Y^L\leq Y^L\leq \tilde{m}_1<\frac{1}{(1-q)\gamma}.$$
	On the other hand, by the concavity of $F_{\Lambda}$ in $(y,z^\perp)$, it implies
	\begin{align}
	F_{\Lambda}(Y^L,Z^{\perp,L})&=F_{\Lambda}\big(\theta Y^U+(1-\theta)\frac{\delta_\theta Y}{1-\theta},\theta Z^{\perp,U}+(1-\theta)\frac{\delta_\theta Z^{\perp}}{1-\theta}\big)\nonumber\\
	&\geq \theta F_{\Lambda}\big( Y^U, Z^{\perp,U})+	(1-\theta)F_{\Lambda}(\frac{\delta_\theta Y}{1-\theta},\frac{\delta_\theta Z^{\perp}}{1-\theta}\big).	\label{deltaF}
	\end{align}

	Using It\^{o}'s formula to $\exp_q(- \frac{\gamma}{1-\theta}\delta_\theta Y)$, it obtains that
	\begin{align*}
	&d\exp_q(- \frac{\gamma}{1-\theta}\delta_\theta Y_s)\\
	=&-\exp_q^q(- \frac{\gamma}{1-\theta}\delta_\theta Y_s)\frac{\gamma}{1-\theta}(\delta_\theta Z^{\perp}_s\cdot dW_s^\perp+\delta_\theta Z_s\cdot dW_s^{-\lambda})\\
	&+\exp_q^q(- \frac{\gamma}{1-\theta}\delta_\theta Y_s)\frac{\gamma}{1-\theta}\delta_\theta F_{\Lambda}(s)ds\\
	&+\frac{q}{2}\frac{\gamma^2}{(1-\theta)^2}\exp_q^{2q-1}(- \frac{\gamma}{1-\theta}\delta_\theta Y_s)(|\delta_\theta Z^{\perp}_s|^2+|\delta_\theta Z_s|^2)ds\\
	\geq& -\exp_q^q(- \frac{\gamma}{1-\theta}\delta_\theta Y_s)\frac{\gamma}{1-\theta}(\delta_\theta Z^{\perp}_s\cdot dW_s^\perp+\delta_\theta Z_s\cdot dW_s^{-\lambda})\\
	&+\frac{q}{2}\frac{\gamma^2}{(1-\theta)^2}\exp_q^{2q-1}(- \frac{\gamma}{1-\theta}\delta_\theta Y_s)|\delta_\theta Z_s|^2ds,
	\end{align*}
	where the last inequality is derived from \eqref{deltaF} and the definition of $F_{\Lambda}$.
	
	For each $\theta\in (0,1)$, since $\exp_q^q(- \frac{\gamma}{1-\theta}\delta_\theta Y_s)$ is bounded, and
	$\delta_\theta Z^{\perp}$ and $\delta_\theta Z$ is square integrable, then the stochastic integral term is a martingale under $\mathbb{Q}^{min}$. Taking the conditional expectation under $\mathbb{Q}^{min}$, we get that
	$$\exp_q(- \frac{\gamma}{1-\theta}\delta_\theta Y_t)\leq \mathbb{E}_{\mathbb{Q}^{min}}\left[\exp_q(- \gamma\xi) ~|\mathcal{F}_t\right],~~t\in[0,T].$$
	Therefore, for each $t\in[0,T]$,
	\begin{align*}Y^L_t-\theta Y^U_t&=\delta_\theta Y_t\\
	&\geq -(1-\theta)\frac{1}{\gamma}\ln_q\mathbb{E}_{\mathbb{Q}^{min}}\left[\exp_q(-\gamma\xi) ~|\mathcal{F}_t\right]\\
	&\geq (1-\theta) \tilde{m}_2.\end{align*}
	Sending $\theta$ to $1$, we get $Y^L-Y^U \geq 0$, which
	gives the uniqueness result.
	
	Case (ii): $q>1$. For any $\theta\in(0,1)$, setting
	\begin{align*}
	\tilde\delta_\theta Y=Y^U-\theta Y^L,
	\tilde\delta_\theta Z=Z^U-\theta Z^L, \tilde\delta_\theta Z^{\perp}=Z^{\perp,U}-\theta Z^{\perp,L}.
	\end{align*}Similarly, we get
	$$\frac{\tilde\delta_\theta Y}{1-\theta}=\frac{\theta}{1-\theta}(Y^U-Y^L)+Y^U\geq Y^U\geq \tilde{m}_2>\frac{1}{(1-q)\gamma}$$
	and
	\begin{align*}\tilde\delta_\theta F_{\Lambda}(s)&= F_{\Lambda}\big( Y^U, Z^{\perp,U})-\theta F_{\Lambda}(Y^L,Z^{\perp,L})\\
	&\geq (1-\theta)F_{\Lambda}(\frac{\tilde\delta_\theta Y}{1-\theta},\frac{\tilde\delta_\theta Z^{\perp}}{1-\theta}\big).
	\end{align*}
	
	Applying It\^{o}'s formula to $\ln \mu( \frac{\tilde\delta_\theta Y}{1-\theta})$, we get
	\begin{align*}
	&d\ln \mu( \frac{\tilde\delta_\theta Y_s}{1-\theta})\\
	=&\frac{(q-1)\gamma }{q(1-\theta)}\mu^{-1}( \frac{\tilde\delta_\theta Y_s}{1-\theta})d\tilde\delta_\theta Y_s-\frac{1}{2}\frac{(q-1)^2\gamma^2}{q^2(1-\theta)^2}\mu^{-2}( \frac{\tilde\delta_\theta Y_s}{1-\theta})\langle d\tilde\delta_\theta Y_s\rangle^2\\
	=&\frac{(q-1)\gamma }{q(1-\theta)}\mu^{-1}( \frac{\tilde\delta_\theta Y_s}{1-\theta})(\tilde\delta_\theta Z_s\cdot dW_s^{-\lambda}+\tilde\delta_\theta Z^{\perp}_s\cdot dW_s^\perp)\\
	&-\frac{(q-1)\gamma }{q(1-\theta)}\mu^{-1}( \frac{\tilde\delta_\theta Y_s}{1-\theta})\tilde\delta_\theta F_{\Lambda}(s)ds\\
	&-\frac{1}{2}\frac{(q-1)^2\gamma^2}{q^2(1-\theta)^2}\mu^{-2}( \frac{\tilde\delta_\theta Y_s}{1-\theta})(|\tilde\delta_\theta Z^{\perp}_s|^2+|\tilde\delta_\theta Z_s|^2)ds\\
	\leq &b_s\cdot dW_s^{-\lambda}-\frac{1}{2}|b_s|^2ds+(q-1)a_s\cdot dW^\perp_s-\frac{1}{2}(q-1)^2|a_s|^2ds\\
	&-\frac{(q-1)\gamma }{q}\mu^{-1}( \frac{\tilde\delta_\theta Y_s}{1-\theta})F_{\Lambda}(\frac{\tilde\delta_\theta Y_s}{1-\theta},\frac{\tilde\delta_\theta Z^{\perp}_s}{1-\theta}\big)ds\\
	=&b_s\cdot dW_s^{-\lambda}-\frac{1}{2}|b_s|^2ds+(q-1)a_s\cdot dW^\perp_s+\frac{1}{2}(q-1)|a_s|^2ds,
	\end{align*}
	where $$b_s=\frac{(q-1)\gamma }{q(1-\theta)}\mu^{-1}( \frac{\tilde\delta_\theta Y_s}{1-\theta})\tilde\delta_\theta Z_s$$ and  $$a_s=\frac{\gamma }{q(1-\theta)}\mu^{-1}( \frac{\tilde\delta_\theta Y_s}{1-\theta})\tilde\delta_\theta Z^\perp_s.$$

	Integrating both sides from $t$ to $T$ and taking the exponent on both sides of the above equality, we get
	$$\mu(\xi)\mathcal{E}(-a\cdot {W}^{\perp})_{t,T}^{q}\leq \mu( \frac{\tilde\delta_\theta Y_t}{1-\theta}) \mathcal{E}\left(b\cdot W^{-\lambda}-a\cdot {W}^{\perp}\right)_{t,T},$$
	which means
	$$\exp_q^{1-q}(-\gamma \xi)\mathcal{E}(-a\cdot {W}^{\perp})_{t,T}^{q}\leq \exp_q^{1-q}(-\gamma \frac{\tilde\delta_\theta Y_t}{1-\theta}) \mathcal{E}\left(b\cdot W^{-\lambda}-a\cdot {W}^{\perp}\right)_{t,T}.$$
	
	For each $\theta\in (0,1)$, due to the boundedness of $\mu^{-1}( \frac{\tilde\delta_\theta Y}{1-\theta})$, and the definitions of $a$ and $b$, $\mathcal{E}\left(b\cdot W^{-\lambda}-a\cdot {W}^{\perp}\right)$ then is a martingale under $\mathbb{Q}^{min}$. Taking the conditional expectation under $\mathbb{Q}^{min}$, and using the reverse H\"{o}lder inequality,  we have, for each $t\in[0,T]$,
	\begin{align*}
	\exp_q^{1-q}(-\gamma \frac{\tilde\delta_\theta Y_t}{1-\theta})&\geq \mathbb{E}_{\mathbb{Q}^{min}}\big[\exp_q^{1-q}(-\gamma \xi)\mathcal{E}(-a\cdot {W}^{\perp})_{t,T}^{q}~|~\mathcal{F}_t\big]\\
	&\geq \mathbb{E}_{\mathbb{Q}^{min}}\big[\mathcal{E}(-a\cdot {W}^{\perp})_{t,T}|~\mathcal{F}_t\big]^q\cdot \mathbb{E}_{\mathbb{Q}^{min}}\big[\exp_q(-\gamma \xi)~|~\mathcal{F}_t\big]^{1-q}.
	\end{align*}
	Finally, we get $\exp_q(-\gamma \frac{\tilde\delta_\theta Y_t}{1-\theta})\geq\mathbb{E}_{\mathbb{Q}^{min}}\big[\exp_q(-\gamma \xi)~|~\mathcal{F}_t\big]$. Therefore, for each  $t\in[0,T]$,
	\begin{align*}Y^U_t-\theta Y^L_t&=\tilde\delta_\theta Y_t\\
	&\leq -(1-\theta)\frac{1}{\gamma}\ln_q\mathbb{E}_{\mathbb{Q}^{min}}\left[\exp_q(-\gamma\xi) ~|\mathcal{F}_t\right]\\&\leq -(1-\theta) \tilde{m}_1.\end{align*}
Sending $\theta$ to $1$, we get $Y^U-Y^L \leq 0$, which gives the uniqueness result.
\hfill$\Box$

\section*{Acknowledgments} The author would like to thank the editor, the associate editor, and two referees for their valuable comments and suggestions which led to a much improved version of the paper.  Part of the work was completed by Dr. Tian during his visit to School of Mathematics, Shandong University. The warm hospitality of Shandong University is gratefully acknowledged. The author thanks Prof. Shengjun Fan for his helpful discussions. 


\bibliography{references}

\end{document}